\documentclass{aa}  
\usepackage{graphicx}
\usepackage{multirow}
\usepackage{xcolor}
\usepackage[caption=false]{subfig}
\usepackage{bm}
\usepackage{txfonts}
\usepackage[colorlinks=true, allcolors=blue]{hyperref}

\usepackage{ulem}

\newcommand{\lowmass}{42}
\newcommand{\lowmassperc}{49\%}

\begin{document}

   \title{Spectroscopic substellar initial mass function of NGC 2244
   \thanks{Based on observations collected at the European Southern Observatory under ESO programme 0102.C-0682(A)}
   }
    \authorrunning{Almendros-Abad et al.}
    
   \author{ V. Almendros-Abad \inst{1}, K. Mu\v{z}i\'c\inst{1,2}, H. Bouy\inst{3}, A. Bayo\inst{4,5}, A. Scholz\inst{6}, K. Peña Ramírez\inst{7}, A. Moitinho\inst{1}, K. Kubiak\inst{1}, R. Schöedel\inst{8}, R. Barač\inst{9}, P. Brčić\inst{9}, J. Ascenso\inst{2} \and R. Jayawardhana\inst{10}
          }

   \institute{CENTRA, Faculdade de Ci\^{e}ncias, Universidade de Lisboa, Ed. C8, Campo Grande, P-1749-016 Lisboa, Portugal\\
              \email{valmendros@sim.ul.pt}
        \and
            Faculdade de Engenharia, Universidade do Porto, Rua Dr. Roberto Frias, 4200-465 Porto, Portugal
        \and
           Laboratoire d’Astrophysique de Bordeaux, Univ. Bordeaux, CNRS, B18N, All\'ee Geoffroy Saint-Hillaire, 33615 Pessac, France
        \and
            European Southern Observatory, Karl-Schwarzschild-Strasse 2, 85748 Garching bei München, Germany     
        \and
            Instituto de F\'{\i}sica y Astronom\'{\i}a, Universidad de Valpara\'{\i}so, Chile
        \and
            SUPA, School of Physics \& Astronomy, University of St Andrews, North Haugh, St Andrews, KY16 9SS, United Kingdom
        \and
            Centro de Astronomía (CITEVA), Universidad de Antofagasta, Av. Angamos 601, Antofagasta, Chile
        \and
            Instituto de Astrofísica de Andalucía (CSIC), Glorieta de la Astronomía s/n, 18008 Granada, Spain
        \and
            Odjel za fiziku, Prirodoslovno-matematički fakultet, University of Split, Ruđera Boškovića 33, 21000 Split, Croatia
        \and 
            Department of Astronomy, Cornell University, Ithaca, NY 14853, USA
            }

   \date{Received; accepted}

  \abstract
   {The dominant formation channel of brown dwarfs (BDs) is not well constrained yet and a promising way to discriminate between scenarios consists of testing environment-dependent efficiency in forming BDs. So far, the outcome of star formation, studied through the initial mass function, has been found to be very similar in all clusters that have been studied.}
   {We aim at characterizing the low-mass (sub)stellar population of the central portion (2.4 pc$^2$) of the $\sim$2 Myr old cluster NGC 2244 using near infrared spectroscopy. By studying this cluster, characterized by a low stellar density and numerous OB stars, we aim at exploring the effect that OB stars may have on the production of BDs.}
   {We obtain near infrared $HK$ spectroscopy of 85 faint candidate members of NGC 2244. We derive the spectral type and extinction by comparison with spectral templates. We evaluate cluster membership using three gravity-sensitive spectral indices based on the shape of the $H$-band. Furthermore, we evaluate the infrared excess from \textit{Spitzer} of all the candidate members of the cluster. Finally, we estimate the mass of all the candidate members of the cluster and derive the initial mass function, star-to-BD number ratio and disk fraction.}
   {The initial mass function is well represented by a power law ($dN/dM \propto M^{-\alpha}$) below 0.4~$M_\odot$, with a slope $\alpha$ = 0.7-1.1 depending on the fitted mass range. We calculate a star-to-BD number ratio of 2.2-2.8. We find the low-mass population of NGC 2244 to be consistent with nearby star-forming regions, although it is at the high-end of BD production. We find BDs in NGC 2244 to be on average closer to OB stars than to low-mass stars, which could potentially be the first evidence of OB stars affecting the formation of BDs. We find a disk fraction of all the members with spectral type later than K0 of 39$\pm$9\% which is lower than typical values found in nearby star-forming regions of similar ages.}
   {}


   \maketitle
%

\section{Introduction}

Brown dwarfs (BDs), objects with masses below 0.075 $M_\odot$, populate the mass spectrum between the realms of stars and planets, being mainly characterized by the lack of hydrogen burning in their cores. As is the case for stars, most BDs are born in clusters and associations, making those perfect laboratories where to study the formation of stars and BDs. One of the most studied observable outcomes of the star formation process is the initial mass function (IMF), which represents the mass distribution at birth. Since the pilot work of \citet{salpeter55}, the high mass side of the IMF has been found to be reproduced by a single power-law with the form $dN/dM \propto M^{-\alpha}$, with $\alpha\sim$ 2.35, except for some notable exceptions \citep{bayo11,stolte06,muzic17,hosek19}. 

The IMF flattens below 1~$M_\odot$ with a turnover at 0.1-0.7 $M_\odot$ \citep{chabrier03}. The IMF on the low-mass side can also be well reproduced by a single power-law. In nearby star-forming regions (SFRs) the low-mass side $\alpha$ has been found to range between 0.5-1 \citep{luhman07,bayo11,pena12,alvesdeoliveira12,lodieu13,scholz13,muzic15,suarez19}. At the same time, it has been found that 2-5 stars are born for every BD in a single SFR \citep{muzic19}. The wide range of values found reflects the overall contribution of the uncertainties of all parameters involved in the derivation of the slope of the IMF and the star-to-BD number ratio, rather than variations between clusters. Based on this result and the comparison of other properties of BDs and stars (e.g. kinematic and spatial distributions, protoplanetary disk properties and multiplicity) it is generally accepted that at least most of the BDs with higher masses form like stars \citep{luhman12}, from the turbulent fragmentation of the molecular cloud \citep{padoan04}. However, due to the intrinsic faintness of BDs, most of the environments where the substellar content has been studied in detail are located nearby and are characterized by relatively loose groups of low-mass stars, with no or very few massive stars. The current BD formation theories predict that the production of BDs could be affected by the environment where they are born. The production of BDs could be increased in environments with high gas and/or stellar densities \citep{bonnell08,jones18}, that may also favor BD formation through ejections \citep{bate12}. The presence of massive stars can also increase the efficiency of BD formation through photoevaporation of the outer layers of a forming protostar \citep{whitworth04} or by fragmentation in their massive disks \citep{stamatellos11,vorobyov13}.

In order to probe environments whose properties in terms of stellar density and number of OB stars are more extreme than that found in nearby SFRs, we have a started a program to study a sample of massive young clusters. In \citet{muzic17} we studied the core of RCW 38, a young (1 Myr), embedded cluster located at a distance of 1.7 kpc. RCW 38 is characterized both by high stellar densities (core stellar surface density $\Sigma$~$\sim$2500 pc$^{-2}$, twice as dense as the ONC and more than ten times denser than NGC 1333) and by a substantial population of massive OB stars \citep[$\sim$60 OB stars,][]{wolk08,winston11}. We found the IMF of the cluster to be consistent with nearby star forming regions ($\alpha$~$\sim$0.7 for the range 0.02-0.5 $M_\odot$). In \citet{muzic19} we studied the core of NGC 2244, a young \citep[2 Myr,][]{hensberge00,bell13,wareing18} cluster located at a distance of 1.5 kpc. It is characterized by a low stellar density (core stellar surface density $\Sigma\sim$30 pc$^{-2}$, similar to the sparse nearby SFRs) and a rich population of OB stars \citep[$>$70 massive stars,][]{sfh_rosette}. These OB stars are presumed to be responsible for the evacuation of the gas in the central region of the cluster \citep{sfh_rosette}, and may have also eroded the protoplanetary  disks around stars in their vicinity \citep{balog07}. We found the IMF to be on the high-end of BD production while still consistent with nearby SFRs ($\alpha$~$\sim$1 for the range 0.02-0.5 $M_\odot$). The membership analysis in \citet{muzic19} was carried out in a statistical way, so in order to confirm those results, we need to perform a spectroscopic follow-up. Spectroscopy is a fundamental step in IMF studies, where the membership of cluster candidate members can be confirmed, removing one of the sources for uncertainty in the derivation of the IMF. Thanks to the low extinction and distance to NGC 2244, spectroscopy of candidate members is feasible into the substellar regime with current facilities.

In this work, we present a spectroscopic follow-up of 85 NGC 2244 faint candidate members ($\sim$0.03-0.4 $M_\odot$). In Sect.~\ref{dataset} we present the spectroscopic observations and data reduction, a new reduction of \textit{Spitzer} mid-infrared data and a compilation of all the known stellar members of the core region of the cluster. In Sect.~\ref{analysis} we derive the spectroscopic properties of the candidate members with low-mass signatures and evaluate the youth of the cluster candidate members using gravity-sensitive spectral indices. We also estimate the infrared excess and derive the masses of all the members of the cluster. In Sect.~\ref{imf} we derive the IMF and star-to-BD number ratio and evaluate the disk fraction into the BD regime. In Sect.~\ref{summary} we discuss the effect of massive stars may have had in the formation of BDs in NGC 2244.

\section{Data set} 
\label{dataset}

\begin{figure*}[hbt!]
    \centering
    \includegraphics[width=\textwidth]{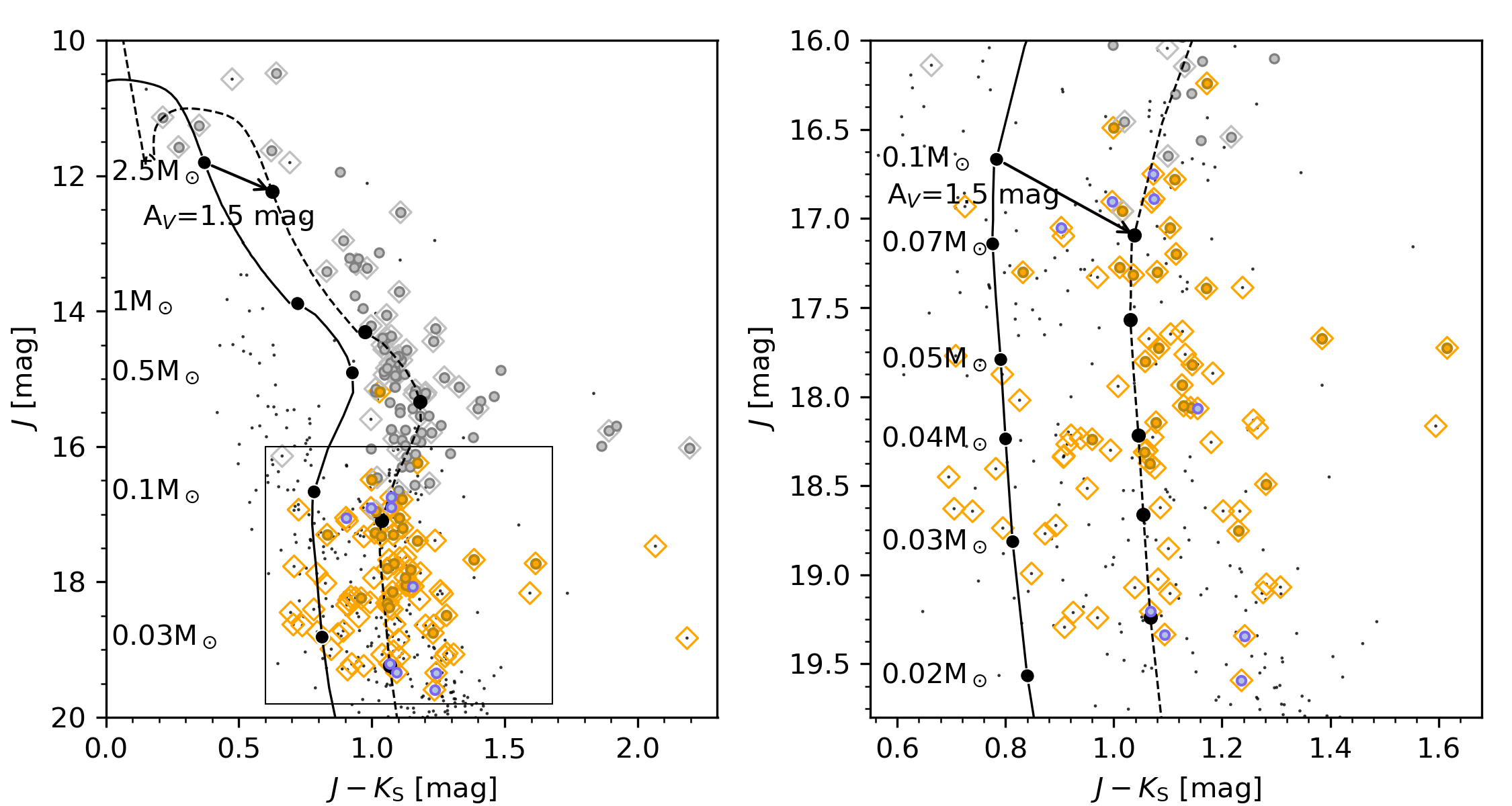}
    \caption{Color magnitude diagram (CMD) of the core of NGC2244. We show the Flamingos-2 catalog \citep{muzic19} with black dots, objects observed with KMOS are represented by orange empty diamonds, KMOS members as solid orange circles, KMOS possible members as blue solid circles, VIMOS observed objects as empty grey diamonds, and members from VIMOS, \citet{muzic22} and \citet{meng17} as grey circles. The black solid line represents the 2 Myr isochrone shifted to the distance of 1500 pc, and the dashed line represents the same isochrone reddened by the mean extinction toward NGC 2244 \citep[i.e. $A\mathrm{_V}$=1.5 mag,][]{muzic19}. The isochrone shown comes from the PARSEC stellar models \citep{bressan12} down to 1 $M_\odot$, and BT-Settl below 1$M_\odot$ \citep{baraffe15}, i.e. isochrone 1 in \citet{muzic19}. The right panel represents the inset of the CMD to the region populated by the KMOS observations. The black arrow represents the effect of $A\mathrm{_V}$=1.5 mag.}
    \label{fig:cmd}
\end{figure*}

\subsection{Near infrared spectroscopy}

\citet{muzic19} analyzed the IMF of the core region of NGC 2244 into the BD regime using deep photometric observations. These observations were taken with the Flamingos-2 near-infrared camera at the Gemini-South telescope in the $JHK$ bands on a field of $\sim$2.4 pc$^2$ (black dots in Fig.~\ref{fig:cmd}, we refer to these observations as the Flamingos-2 catalog hereafter). We use the same ID notation to refer to the candidate members of NGC 2244\footnote{The Flamingos-2 catalog is available in electronic form at the CDS: \url{https://cdsarc.cds.unistra.fr/viz-bin/cat/J/ApJ/881/79}}. The 90\% completeness limit of the Flamingos-2 catalog is $J$=21.7 mag. Here we present the spectroscopic follow-up of 135 sources too faint to have proper motions from Gaia ($J>$16.5 mag, see \citealt{muzic22}) down to $J$~$\sim$19.8 mag (limited by the feasibility of the observations).

\subsubsection{Observations}

The observations were carried out using the K-band Multi Object Spectrograph \citep[KMOS,][]{sharpies13}, located at the Very Large Telescope (VLT), under program 0102.C-0682(A) (PI: K. Muzic). KMOS performs integral field spectroscopy and has 24 arms, each with a squared field of view (FoV) of $2.8''\times2.8''$. The total FoV of the instrument is $7.2'$ in diameter. KMOS was operated in service mode using a nod to sky ABA configuration, where A is the target and B is the sky. Between each exposure a small dithering was performed. In order to optimize the integration time, we divided the observations in two groups according to their brightness: bright ($J <18.5\,$mag) and faint ($J > 18.5\,$mag). For the bright set of observations, 10 exposures with 90s detector integration time (DIT) each were obtained, whereas for the faint set, 20 exposures with DIT = 120\,s were taken. The spectra were obtained using the $HK$ filter ($\sim$1.5-2.4 $\mu$m), which has a mean spectral resolving power of $\sim$1800. The sources were randomly selected from the complete Flamingos-2 catalog  with $J$=16.5-19.5 mag to maximize the number of arms allocated in each observation block. This process was optimized using the KMOS ARM Allocator (KARMA) software. In this process we also allocated one brighter source (\#770) to a free arm. Together with the target spectra, all the calibration frames needed for a successful data reduction were also obtained: dark, flat, arc-lamp, standard star and telluric standard stars.

\subsubsection{Data reduction}
\label{data_reduction}

The data reduction was performed using the KMOS pipeline \citep[SPARK,][]{davies13} in the ESO Reflex automated data reduction environment \citep{freudling13}. The pipeline produces calibrated 3D cubes, by performing flat field correction, wavelength calibration, sky subtraction and telluric correction \citep[using \textit{Molecfit},][]{smette15}.  However, we noticed that the quality of the subtraction of the OH sky lines (telluric emission features) in the data products was poor, especially for the cases with lower signal-to-noise ratio ($S/N$). We found \textit{SkyCorr} \citep{noll14} to improve the OH sky line residuals in the $H$-band and the signal at the end of the $K$-band and in the $\sim$2-2.05 $\mu m$ region where there is a telluric H$_2$O absorption band. Therefore, we used the instrument's pipeline to perform the entire data reduction except for sky subtraction. Since the objects were not discernible in the data cubes without sky subtraction, we used the pipeline sky subtracted data cubes to define the spectrum extraction mask. We fitted a 2-D Gaussian to each individual exposure of the pipeline sky-subtracted data cubes. The 2-D Gaussian was used to define a mask with the weights of the extraction at each spaxel. The mask was applied to the non-sky subtracted data cubes, as well as to the corresponding dithered sky exposures. Then, we performed the sky subtraction using \textit{SkyCorr}, the input being the extracted object and sky 1D spectra. We found that \textit{SkyCorr} performed better when the $H$ and $K$-band spectra are treated separately, removing the region in between the bands (1.8-1.95 $\mu m$) that is dominated by telluric absorption. Finally, we combined all the exposures of each object with a median approach. The noise of the spectra at each wavelength was derived as:
    \begin{equation}
        N=\frac{\sigma}{\sqrt{n}},
    \end{equation}
where N is the noise, $\sigma$ is the standard deviation of all the exposures at each wavelength, and n is the number of exposures.

We inspected whether the signal obtained in each spectrum is enough to perform any analysis. An initial visual inspection of the spectra confirmed that 50 out of the original 135 spectra had a low $S/N$ ($S/N$$<$4 in $H$-band) which prevented any meaningful analysis, and were therefore excluded from the data set. Based on the brightness of the remaining sources and the distance and average extinction of NGC 2244 \citep{muzic22}, the masses of the observed candidate members should roughly correspond to 0.03 - 0.4 $M_\odot$.

\subsection{Mid-infrared photometry}

We performed a new reduction of the available \textit{Spitzer} data of the NGC 2244 region. We retrieved all the images obtained with IRAC \citep{fazio04} on-board the \textit{Spitzer Space Telescope} \citep{werner04} from the Spitzer Heritage Archive within a radius of 1.5\degr\, centered on NGC2244. Table~\ref{tab:programs_spitzer} gives the list of program ID, principal investigator and associated exposure times and number of frames (i.e. \# of BCD). Our reduction began from the S18.25.0 pipeline-processed artifact Corrected Basic Calibrated Data (CBCD). We then combined these into mosaics using the recommended  version 18 of MOPEX (MOsaicker and Point source EXtractor) provided by the \textit{Spitzer} Science Center using the standard parameters (see the MOPEX User’s Guide for details on the data reduction). We chose to process the short 0.4s and long 10.4s frames separately and to produce a mosaic for each dataset. This allowed us to nicely cover the largest possible dynamical range. The presence of a bright extended nebulosity compromises the detection of sources and measurement of their photometry as the background estimations implemented in SExtractor \citep{bertin96} and APEX (the photometry package that is part of MOPEX) are not optimised to deal with such variable extended emission. We therefore applied the nebulosity filter described in Bertin et al. (in prep) to the pipeline produced mosaics. Sources were then detected using SExtractor, and their aperture photometry was then extracted using APEX. Apertures of 1\farcs2, 1\farcs8 and 2\farcs4 radii and a sky annulus between 14\farcs4 and 24\farcs0 were used to measure the photometry in the four IRAC bands. Finally, the recommended aperture corrections were applied to the fluxes.

Finally, we cross-matched the extracted IRAC catalog with the Flamingos2 catalog. We visually checked that the correspondence between the NIR and mid-infrared sources is correct. In the few cases were the IRAC photometry did not correspond to a single object in the Flamingos-2 catalog we rejected their IRAC photometry.

\begin{table}
\centering
\caption{Spitzer/IRAC observations of NGC2244}
\begin{tabular}{cccc}
\hline
Program ID  &    P.I.  &  EXPTIME (s)   &  \# of BCD    \\
\hline
30726  &   J. Bouwman  &  0.4, 10.4   & 288   \\
3394   &   I. Bonnell   &    0.4, 10.4   & 530      \\
37    &    G. Fazio  &  0.4, 10.4   & 291      \\
40359  &   G. Rieke   &   0.4, 10.4   & 1404     \\
61071   &  B. Whitney  & 0.4, 10.4  &  3188       \\
61073   &  B. Whitney  & 0.4, 10.4  &  2         \\
\hline\end{tabular}
\label{tab:programs_spitzer}
\end{table}

\subsection{Previous members}
\label{memb_high}

We compiled all the previously reported candidate members encompassing the core of NGC224, i.e. the region covered by the Flamingos-2 catalog. \citet{muzic22} built a catalog of probable members of NGC 2244. The members were obtained by applying the Probabilistic Random Forest algorithm using as input optical to  near infrared (NIR) photometry as well as proper motions from \textit{Gaia} EDR3. 85 of these candidate members are in the Flamingos-2 catalog. \citet{muzic22} also presented optical spectroscopy of 501 candidate members of NGC 2244. The objects were observed using VIMOS/VLT. The spectral type (SpT) and visual extinction ($A_{\mathrm{V}}$) were derived for the 327 objects that presented late SpT spectral shape by comparison with spectral templates. Strong fringing in the VIMOS spectra longwards of 8000 $\AA$ prevented performing an analysis of gravity-sensitive absorption lines in that range, leaving H$\alpha$ as the only signature of youth. The equivalent width of H$\alpha$ was measured and compared with accretion thresholds defined in the literature \citep{barrado03,fang09}. If a source had H$\alpha$ emission above these thresholds, this emission is assumed to be associated with accretion, which is a signature of youth. But if the H$\alpha$ emission is below this threshold, then this emission is also consistent with chromospheric activity and hence not a definite signature of youth. We found 11 candidate members with accretion-related VIMOS H$\alpha$ emission, 9 of which are in the \citet{muzic22} member list. The remaining two candidate members are below or close to the magnitude completeness limit of \citet{muzic22}, which is most probably why they were not selected as members there.

\citet{meng17} presented a catalog of probable members of NGC 2244 based on X-ray data \citep{wang08} and infrared excess from Spitzer \citep{balog07}. There are 84 sources from this member list in the Flamingos-2 catalog, 23 of which are not present either in \citet{muzic22} or have accretion-related H$\alpha$ emission in their VIMOS spectra. Most of these 23 sources are below or close to the magnitude completeness limit of \citet{muzic22} as well. Eight of these sources have a VIMOS spectrum, seven of which present a spectroscopic shape typical of low-mass stars and BDs. The remaining source (\#734) is brighter and its spectrum would not show low-mass signatures. We visually examined the colors, proper motions and parallaxes of the 23 sources and only found three to be inconsistent with membership to the cluster (\#146, \#623, \#754). The remaining 20 sources are considered candidate members of the cluster. Overall, we find 107 candidate members of the cluster, all of them are shown as grey circles in the color-magnitude diagram in Fig.~\ref{fig:cmd}.

\section{Analysis}
\label{analysis}

In this section we present the membership assessment (Sect.~\ref{analysis_youth} and \ref{analysis_census}) and mass derivation of all the NGC 2244 candidate members. We also corrected the census of cluster members for the fact that we did not observe with KMOS all the sources in the faint regime (see Sect.~\ref{analysis_kmos_stat}).

\subsection{Near infrared spectroscopy}
\label{analysis_kmos}

\begin{figure}[hbt!]
    \centering
    \includegraphics[width=\textwidth*10/21]{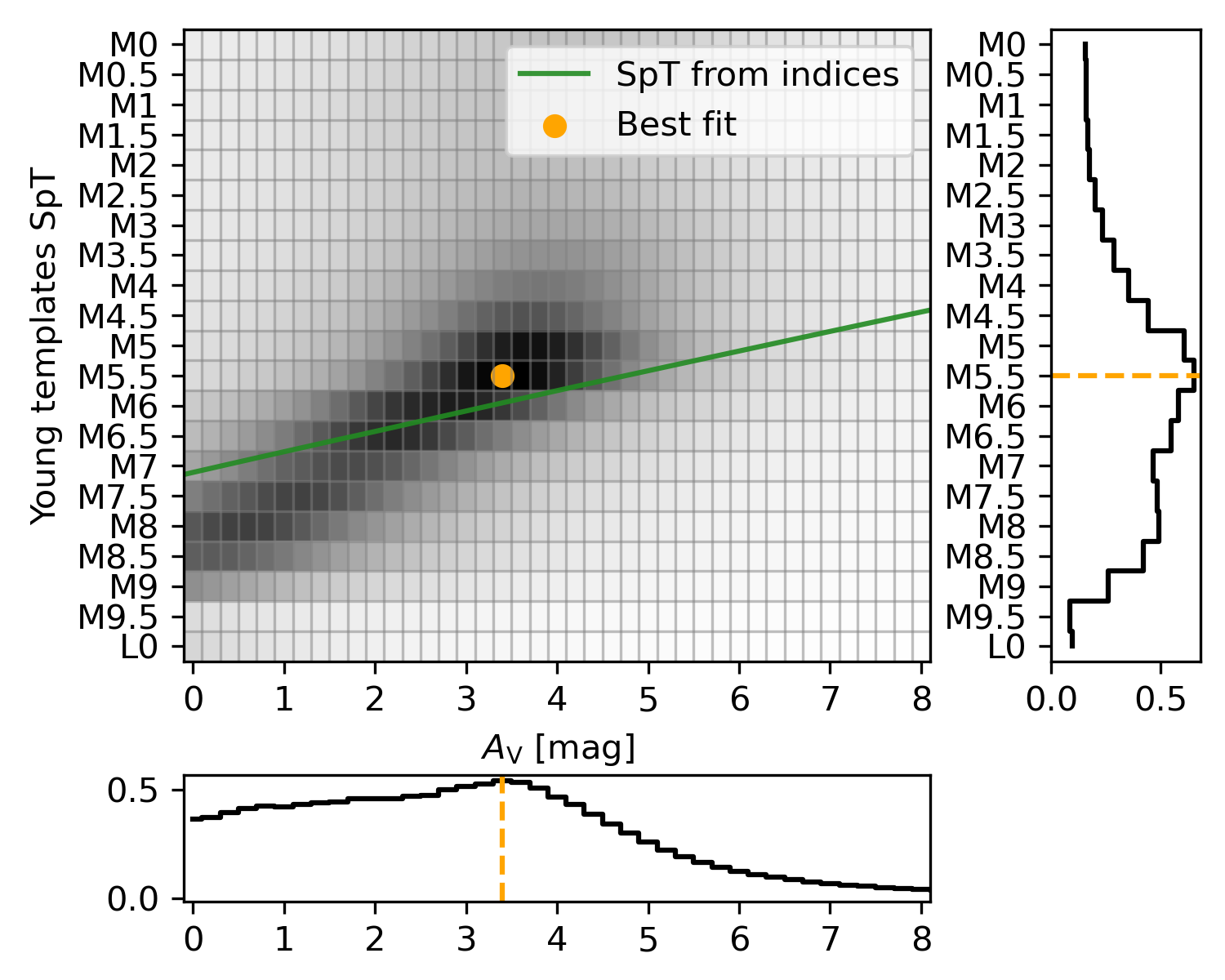}
    \caption{Probability map of SpT and $A\mathrm{_V}$ from comparison with spectral templates. Darker regions represent lower values of $\chi ^2$ from the comparison with the young templates for object \#40. We show the best fit SpT and $A_{\mathrm{V}}$ in orange. The green line represents the SpT derived from a combination of various spectral indices (see Sect.~\ref{analysis_indices}) for the entire $A_{\mathrm{V}}$ grid. At the right and bottom panels we also show the $\chi ^2$ histogram of SpT and $A_{\mathrm{V}}$.}
    \label{fig:probmap}
\end{figure}

In this section we present the analysis of the KMOS NIR spectra. First, we visually inspect all the spectra looking for well-known features present in the NIR spectrum of low-mass stars and BDs. We derive the SpT and extinction of the sources that present these features by comparison with spectral templates. We validate the SpT measurement by comparing it with the SpT coming from spectral indices. Finally, we assess the membership to the cluster of these candidate members using three gravity-sensitive spectral indices. These indices indicate whether the studied objects have a low surface gravity and are therefore young.

\subsubsection{Visual inspection}
\label{analysis_visual}

\begin{figure*}[hbt!]
    \centering
    \includegraphics[width=\textwidth]{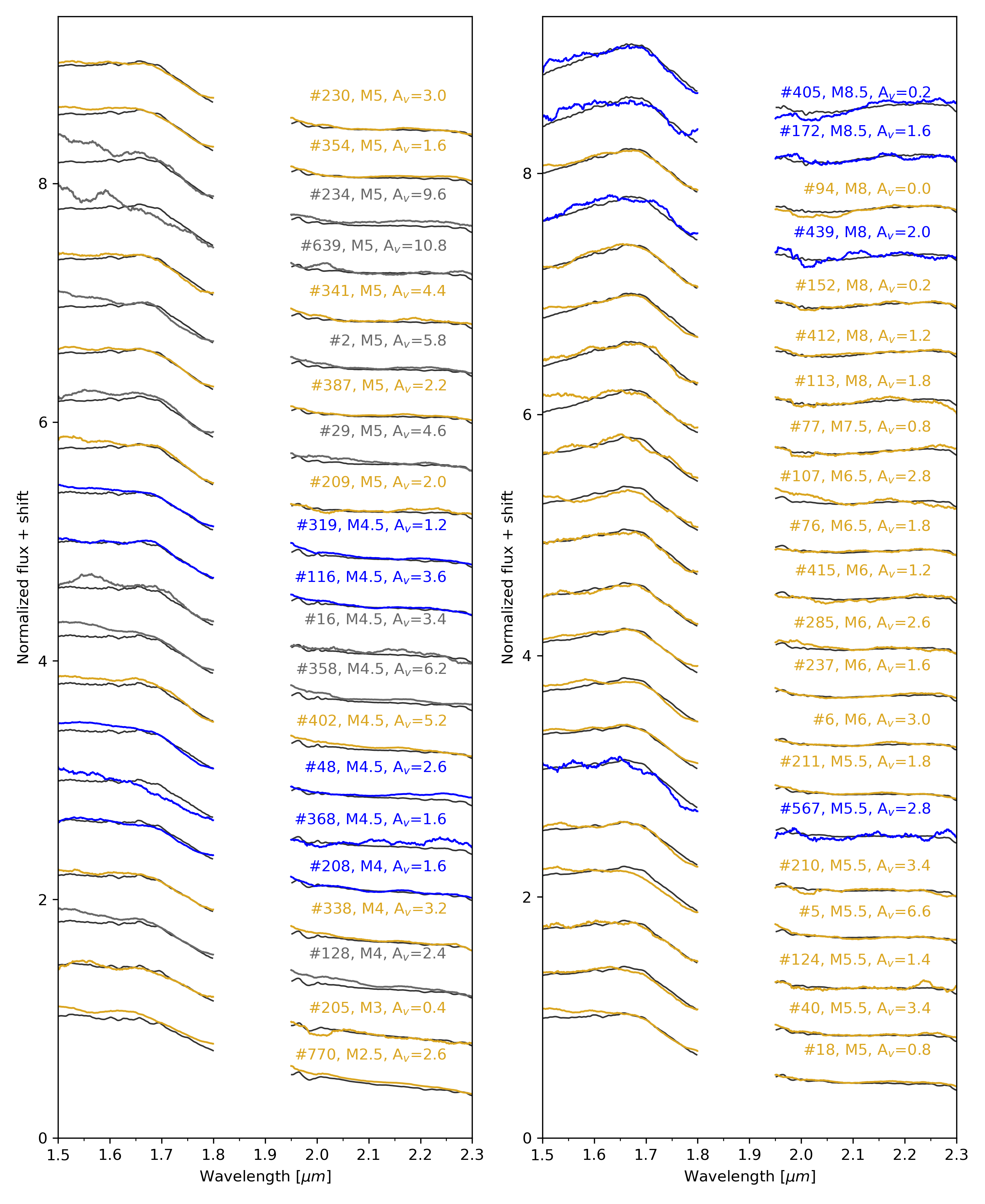}
    \caption{All the KMOS spectra of objects consistent with low-mass (sub)stellar nature, together with the best fit young templates (black) de-reddened by the best fit extinction. The selected SpT and $A_{\mathrm{V}}$ (in magnitude units) are also shown. The color-coding represents the final membership of the targets: orange are members, blue are possible members and grey are non-members.}
    \label{fig:kmos_templates}
\end{figure*}

\begin{figure}[hbt!]
    \centering
    \includegraphics[width=\textwidth*10/21]{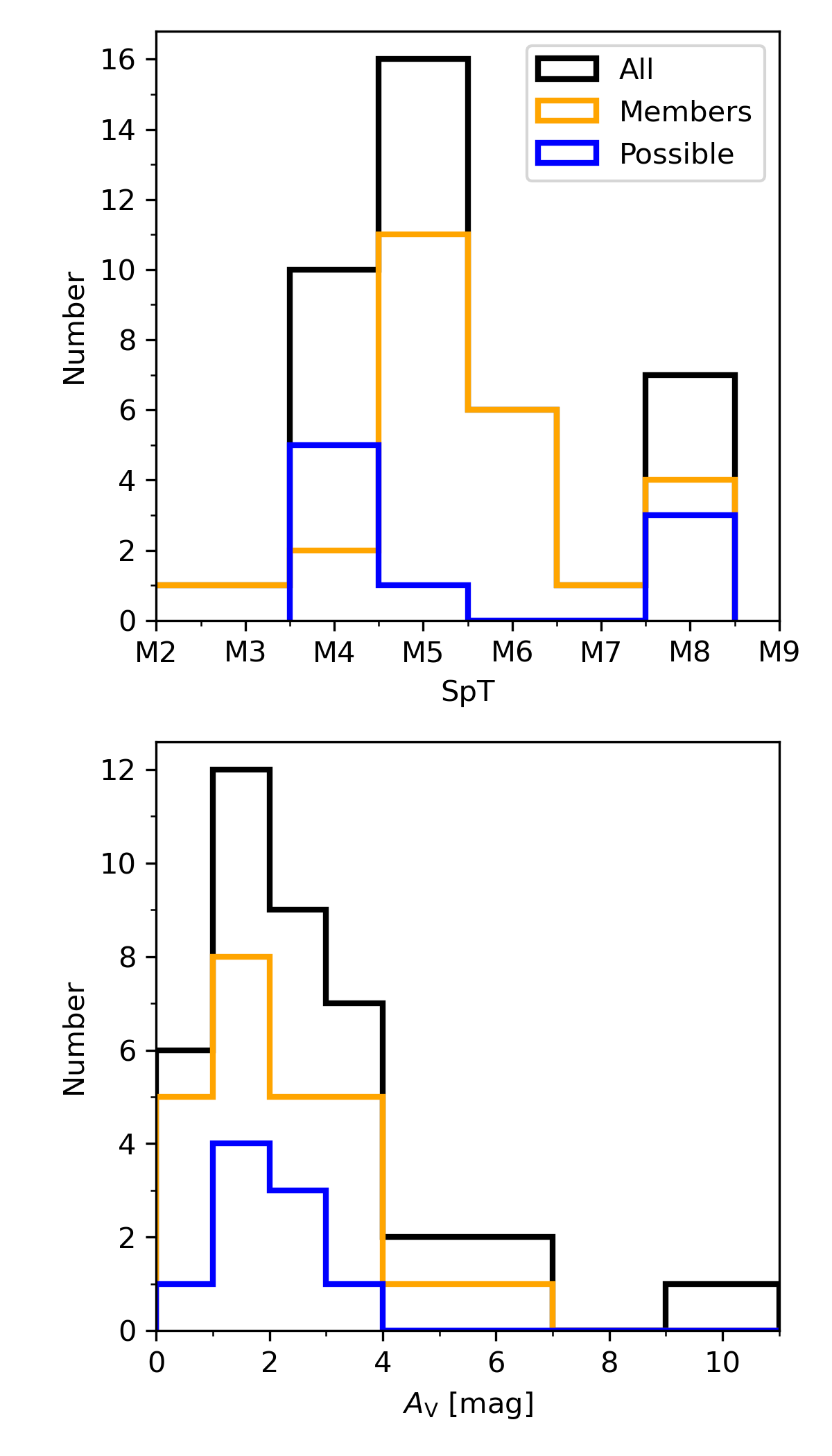}
    \caption{Histogram of SpT and $A_{\mathrm{V}}$ of the objects observed with KMOS derived by direct comparison with spectral templates. We also show the histograms of the members (orange) and possible members (blue).}
    \label{fig:spt_av_visual}
\end{figure}

Based on the expected mass range of the observed candidates ($\sim$0.03 - 0.4 $M_\odot$), the members within the observed candidates should roughly have SpT$\geq$M2. The $HK$ bands spectra of cool dwarfs present distinct characteristics that make them visually identifiable. We therefore perform a visual analysis of the KMOS spectra looking for features characteristic of late SpT objects. The $HK$ bands of cool dwarfs are dominated by broad H$_2$O absorption bands at both red and blue edges of the $H$ and $K$ bands. These features start to dominate at SpT $\sim$M4 and their strength increase with decreasing SpT \citep{cushing05}. The NIR spectrum of cool dwarfs is also shaped by numerous absorption features coming from different species (e.g. CO, NaI, KI, CaI), but they are not well resolved in the KMOS spectra due to their noisy nature. Therefore we are only able to visually inspect the broad shape of the $H$ and $K$ bands. Through visual inspection we find \lowmass~(\lowmassperc~of the targets) to have late SpT spectroscopic shape. The rest of the objects are not discussed further since they present a flat spectrum not in accordance with the cool dwarf shape discussed above. Low-mass members would present a flat spectrum, and therefore be missed in this step, if they are at an earlier evolutionary class (i.e. Class I). However, NGC 2244 has a very low fraction of Class I sources \citep{balog07,poulton08}, so the chance of having observed a Class I source is very low. The contaminants will then most probably be background reddened high-mass stars and extra galactic sources.

\subsubsection{Comparison with spectral templates}
\label{analysis_templates}

For the \lowmass~late SpT candidate members, we derive their SpT and extinction simultaneously by comparison with spectral templates. We use the young ($<$10 Myr) spectral templates from \citet{luhman17}. The templates are defined for half-integer SpTs from M0 down to L0, complemented with L2, L4 and L7 templates. Each object is directly compared with all the spectral templates for a wide range of extinction values: $A_{\mathrm{V}}$=0--12 mag with a step of 0.2 mag. We use the extinction law from \citet{cardelli89} with $R_{\mathrm{V}}$=3.1 \citep{fernandes12} to model the effect of extinction. The object spectra were resampled to match the spectral template wavelength grid. The comparison is made for the $H$ and $K$ bands neglecting the telluric region in between them (1.8-1.95 $\mu m$). All the spectra are normalized at 1.66 $\mu m$ prior to the analysis, and to each comparison we add a grid of wavelengths at which the template is normalized in order to help find the best fit. The goodness of fit of each comparison is evaluated using the reduced $\chi^2$ minimization:

\begin{equation}
    \chi^2=\frac{1}{N-m} \sum_{i=1}^{N} \frac{ (O_i-T_i)^2} {\sigma^2 }
\end{equation}

where $O$ is the object spectrum, $T$ the template spectrum, $\sigma$ is the noise of the observed spectrum, $N$ the number of data points, and $m$ the number of fitted parameters ($m$=3). An example of the output distribution of $\chi^2$ is shown in Fig.~\ref{fig:probmap} (trimmed at L0 for visualization purposes), where darker regions represent lower $\chi^2$. We also show in orange the best fit SpT and extinction, and in green the SpT derived using SpT indices evaluated for the entire extinction grid (see Sect.~\ref{analysis_indices}). Cool dwarfs spectra typically show some degree of degeneracy between the spectral type and the extinction \citep{luhman17}, which is also evident in this case.

For each object we associate the SpT and $A_{\mathrm{V}}$ of the template and extinction configuration that minimizes the $\chi^2$ in the template fitting process. We also performed a visual inspection of all the fits. We noticed that in the case of object \#439, the second best fit (M8) reproduced better the blue edges of both the $H$ and $K$ bands than the best fit (M5.5). The spectrum of this object presents a lower $S/N$ and that makes our methodology less precise. We therefore adopted M8 as the SpT of object \#439. In Fig.~\ref{fig:kmos_templates} we show the KMOS spectra together with the best fit template de-reddened by the best fit extinction except for \#439 as discussed above. The spectra are color-coded by their final membership assessment (see Sect.~\ref{analysis_census}): members (orange), possible members (blue) and non-members (grey). The distribution of the retrieved SpT and $A_{\mathrm{V}}$ values is represented in Fig.~\ref{fig:spt_av_visual}.  The objects in our sample show SpTs between M2.5 and M8.5. There is a drop in low-mass sources with SpT M7, which could be a bias of the observations, or a real feature associated with a known dip in the luminosity function of SFRs between M7 and M8 \citep{dobbie02}. This feature was explained to be possibly caused by the formation of dust in the atmospheres at these temperatures. Most objects present an extinction below 4 mag, and the median extinction is 2.3 mag for the entire low-mass sample, dropping to 2 mag if only the bona-fide members are considered. These values are slightly larger than the previously derived mean extinction of the cluster \citep[$\sim$1.5 mag,][]{muzic19}.

\begin{figure}[hbt!]
    \centering
    \includegraphics[width=\textwidth/21*10]{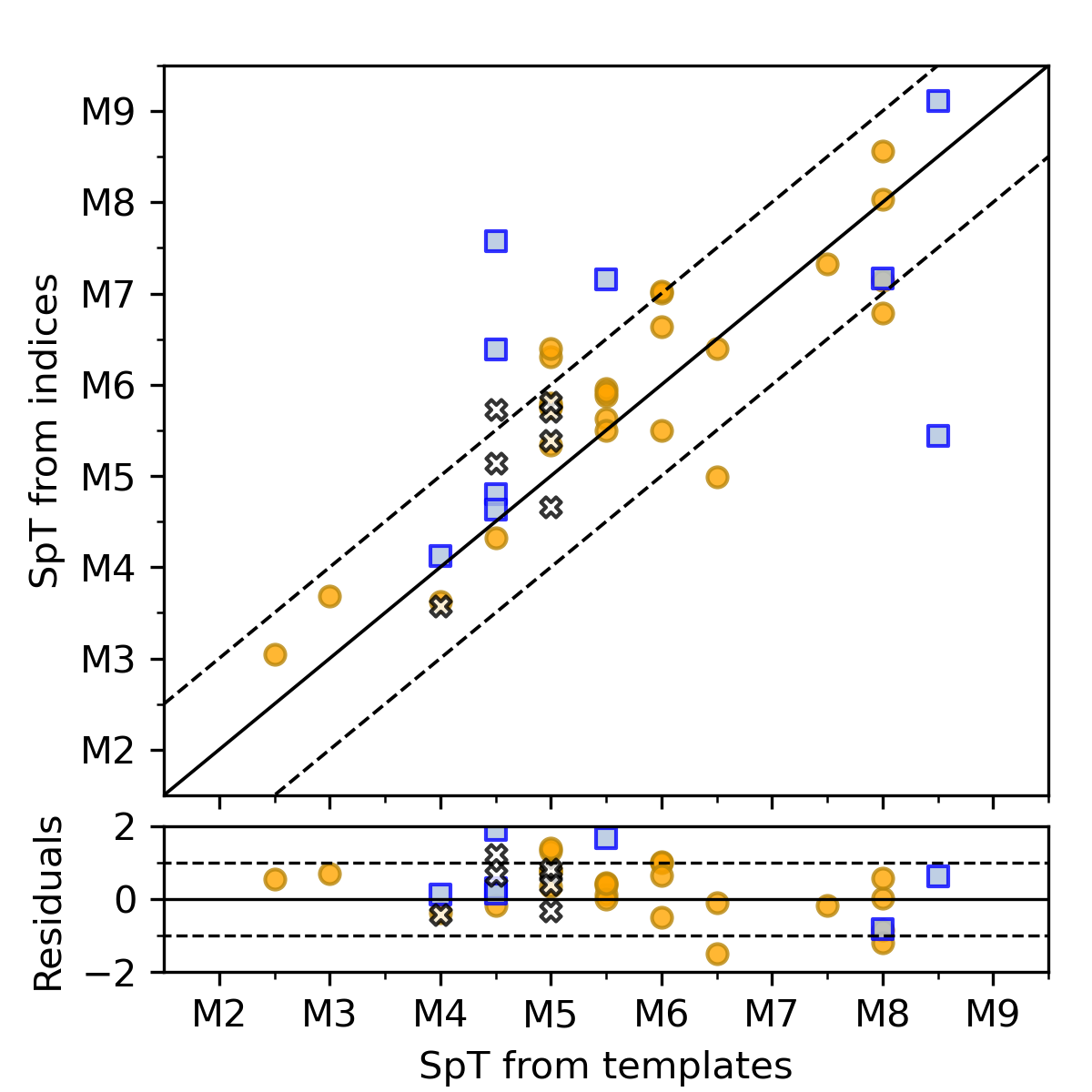}
    \caption{Comparison of the spectral templates best fit SpT (Sect.~\ref{analysis_templates}) with the SpT coming from spectral indices (Sect.~\ref{analysis_census}). The objects are color-coded based on the final membership discussed in Sect.~\ref{analysis_youth}: members (orange circles), possible members (blue squares) and non-members (black crosses). The solid line represents the 1:1 SpT relationship, and the dashed lines represent a $\pm$1 SpT deviation from the 1:1 relationship.}
    \label{fig:compare_spt}
\end{figure}

\subsubsection{Spectral type indices}
\label{analysis_indices}

SpT indices are calibrated to provide a straightforward measurement of the SpT. Their usage is specially interesting when the wavelength range available is constrained to one or two spectral bands, as it is the case here. Using indices for SpT derivation has one main drawback which is that they generally are extinction dependent, therefore extinction needs to be derived by a different method. We de-redden the spectra using the extinction value derived in Sect.~\ref{analysis_templates}. We derive the SpT with the method described in \citet{almendros22} using the indices that fall in the KMOS spectra wavelength range: WK \citep{weights09}, H2O-2 \citep{slesnick04}, sH2O$^{K}$ \citep{testi01} and TLI-K \citep{almendros22}. In Fig.~\ref{fig:compare_spt} we compare the SpT derived here with the value derived in Sect.~\ref{analysis_templates}. We observe that both measurements are in excellent agreement, generally lying within 1 subtype. 

\subsubsection{Youth evaluation}
\label{analysis_youth}

\begin{figure*}[hbt!]
    \centering
    \includegraphics[width=\textwidth]{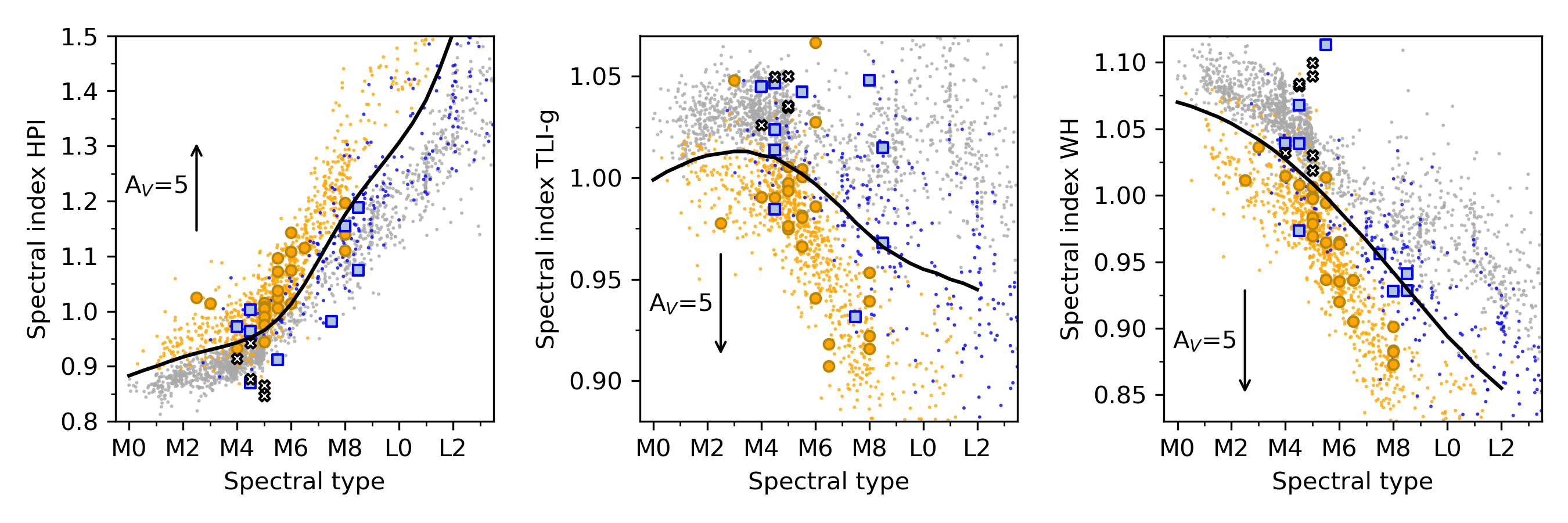}
    \caption{The three used gravity-sensitive indices compared with the SpT. Dots represent the data set from \citet{almendros22} color-coded according to the age class of each object (young in orange, mid-gravity in blue and field in grey). The black solid line represents the decision boundary between young and field objects (see Sect.~\ref{app_decision_boundary}). The objects with KMOS spectra are color- and symbol-coded as in Fig.~\ref{fig:compare_spt}. We also show with a black arrow the effect that an extinction of $A\mathrm{_V}$=5 mag would have on the values of each index.}
    \label{fig:youth_indices}
\end{figure*}

For a late SpT candidate to be confirmed as a member of a young cluster it needs to present some kind of youth indicator in its spectrum. Young objects are still contracting and present a larger radii compared with old objects of the same temperature. As a consequence, young objects possess lower surface gravity than their older counterparts. The NIR spectra of cool dwarfs present several gravity-sensitive features that allow a distinction between field, giant and young low-mass objects \citep{gorlova03,allers13,almendros22}. However, the only feature that can be inspected in our KMOS spectra is the broad band shape of the $H$-band. Low surface gravity objects present a more triangular and sharp shape of the $H$-band than field objects \citep{lucas01}. This feature has been explained by a decrease in H$_2$ collision induced absorption with pressure \citep{kirkpatrick06}. Using machine learning models, \citet{almendros22} showed that the $H$-band peak is indeed the most important feature in the NIR spectra when it comes to surface gravity classification.

There are two gravity-sensitive spectral indices defined to model the change of the $H$-band peak with surface gravity that fall in the KMOS wavelength range: the HPI \citep{scholz12a} and TLI-g \citep{almendros22} indices. Although originally defined for derivation of SpT, \citet{almendros22} showed that the HPI index also presents a gravity-sensitive behavior. The TLI-g index has been shown to be able perform an almost complete separation between young and field cool dwarfs for a wide range of SpT (M0-L3). Additionally, we looked for other SpT indices evaluated in \citet[see their Figs.~C1, C2]{almendros22}, that present a good overall separation of the gravity classes and also fall in the KMOS wavelength range. We find the WH index \citep{weights09} to have a pronounced separation between the gravity classes from very early SpTs (see Fig.~\ref{fig:youth_indices}). We use these three gravity-sensitive indices to perform the membership classification. For the three indices we derive the decision boundary that best represents the separation between the ``young'' and ``field'' sources using the data set presented in \citet[see Sect.~\ref{app_decision_boundary} for further details]{almendros22}.

In Fig.~\ref{fig:youth_indices} we show the three gravity-sensitive indices as a function of SpT for the KMOS sample and the \citet{almendros22} data set together with the derived decision functions. We perform the membership classification as follows: if an object is classified as low-gravity in at least two indices, it is labelled as member (Y); if it is classified as low gravity only in one index, it is labelled as possible member (?); it is labelled as non member otherwise (N, see Table~\ref{tab:tab_results}). With this methodology we aim at keeping all the objects that show any hint of having low surface gravity. In Fig.~\ref{fig:youth_indices} the objects with KMOS spectra are color-coded based on this surface gravity criteria: young members as orange circles, possible members as blue squares and non members objects as empty black crosses. We observe that four of the sources do not display enough $S/N$ in the KMOS spectrum to rely on the measurement of the spectral indices (\#77, \#405, \#439, \#567). We classified these four sources as possible members. Overall, we find 25 bona-fide members and 10 possible members. The distribution of the extinction of the possible members is not different to that of the bona-fide members, all having Av$<$ 4 mag (see Fig.~\ref{fig:spt_av_visual}). There is therefore no clear evidence based on the reddening to distinguish them from foreground contaminants.

\subsection{Infrared excess}
\label{analysis_excess}

Young objects with protoplanetary disks present excess emission at infrared wavelengths coming from the inner part of the disk that is heated by the central object. This feature can be mimicked by other types of objects \citep[see e.g.][]{gutermuth09}, but if the (sub)stellar nature of the source is confirmed, then the infrared excess is a definite proof of the object's youth. While this feature is a very strong indicator of youth, it will not uncover full populations of young objects, since not all young objects have disks \citep{haisch01}. 

We now classify all the candidate members of NGC 2244 based on their infrared excess. We define the excess in the four IRAC channels from the $K_{\mathrm{S}}$ -- IRAC colors. We first derive the $K_{\mathrm{S}}$ -- IRAC photospheric colors as a function of SpT for a compilation of known diskless members of Cha-I \citep{luhman08,esplin17}, Upper Sco \citep{esplin18}, $\rho$-Ophiuchus \citep{esplin20}, Taurus \citep{esplin19} and Corona Australis \citep{esplin22} SFRs (see Sect.~\ref{app_excess}). We observe that in the four IRAC colors an excess of 0.25 mag separates most ($>$97\%) of the diskless sources (see Sect.~\ref{app_excess}). We therefore define 0.25 mag as the boundary to classify an object as having excess in a certain color. 

\begin{figure*}[hbt!]
    \centering
    \includegraphics[width=\textwidth/10*9]{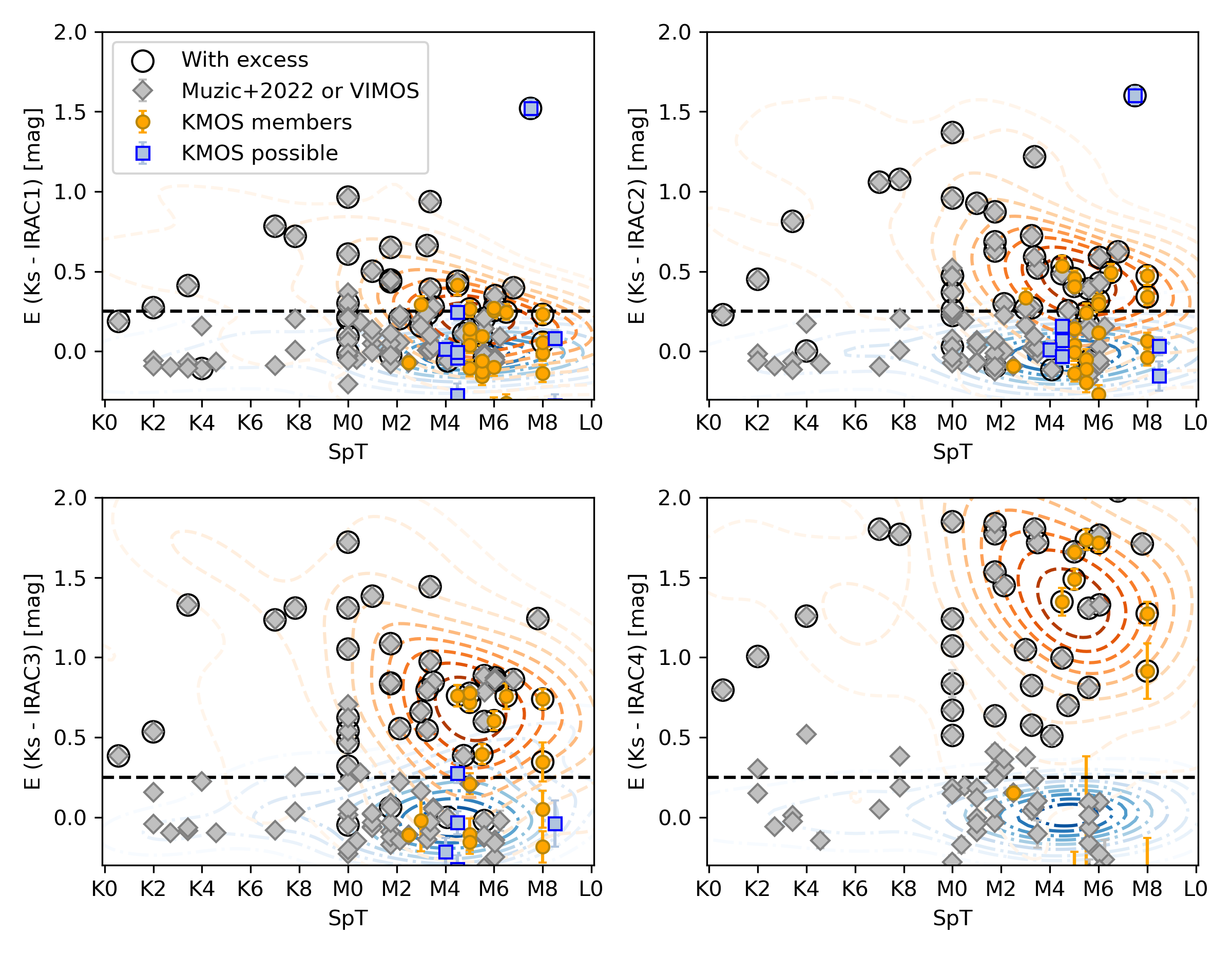}
    \caption{$K_{\mathrm{S}}$ - IRAC color excesses for the candidate members of NGC 2244 with SpT or $T_{\mathrm{eff}}$ measurement. Orange circles represent KMOS bona-fide members, blue squares KMOS possible members and grey diamonds represent VIMOS objects and \citet{muzic22} members. Big black circles represent the objects that we found to have infrared excess. The locus of sources from the literature with ``full'' disks is represented with orange contours, and the locus of sources without disks is represented with blue contours (see Sect.~\ref{app_excess}). Black dashed line represents the 0.25 mag excess boundary.}
    \label{fig:irac_excess}
\end{figure*}

We explore the infrared excess of all the NGC 2244 candidate members that have a SpT measurement, i.e. the candidate members with KMOS or VIMOS spectra. \citet{muzic22} derived the $T_{\mathrm{eff}}$ together with extinction from fitting the spectral energy distribution (SED) using the Virtual Observatory SED Analyzer (VOSA; \citealt{bayo08}). We convert the $T_{\mathrm{eff}}$ to SpT using the \citet{pecaut13} $T_{\mathrm{eff}}$--SpT scale (see Sect.~\ref{analysis_hrd} for justification of the choice of scale). We therefore also evaluate the infrared excess of the \citet{muzic22} candidate members. We classify an object as having excess when it presents excess in at least two colors. We require two colors for the excess classification in order to discard objects that are marginally detected in a single band. With this criteria we may be missing sources with excess only at longer wavelengths (IRAC3 or IRAC4) that are associated with disks at later stages of their evolution such as transitional disks. Additionally, the color excess is affected by the determination of the SpT, which is specially important at later SpTs ($>$M0) where the photospheric colors have a steep relationship with SpT. Therefore, we also visually inspect the SED of all the sources classified as having infrared excess and those that present excess only in IRAC3 or IRAC4. For each source we compare their SED with the BT-Settl atmospheric model \citep{allard12} of the $T\mathrm{_{eff}}$ and $A\mathrm{_V}$ used in Sect.~\ref{analysis_hrd}. We find 6 candidate members to have excess only in IRAC4 (\#33, \#49, \#95, \#197, \#726, \#763), and we classify them as having excess. We also find 4 candidates with color excess to be diskless when inspecting their SEDs (\#205, \#237, \#737, \#739). In Fig.~\ref{fig:irac_excess} we show the color excess as a function of SpT for all the members of the cluster in the four IRAC bands. The locus for sources with ``full'' disks and those that are diskless in the four colors are represented with orange and blue contours respectively. The NGC 2244 candidate members that are classified with infrared excess are highlighted with a big black empty circle.

\subsection{Final census of members}
\label{analysis_census}

\begin{table}
    \caption{Number of members (and possible members) of the core region of NGC 2244 coming from the different member lists. In the diagonal cells we show the total number of members from each member list, and the rest of the cells represent the intersection between them.}
    \begin{center}
        \begin{tabular}{r c c c c}
            \hline\hline
             & M22$^a$ & VIMOS$^b$ & KMOS$^c$ & M17$^d$ \\ 
            \hline
            M22 & 85 & 16 & 2 & 62 \\ 
            VIMOS &  & 24 & 0 & 22 \\ 
            KMOS &  &  & 26 (9) & 4 (1) \\ 
            M17 &  &  &  & 70 (14) \\ 
            \hline 
        \end{tabular}
        \\$^a$Membership in \citet{muzic22}.
        $^b$Objects with VIMOS spectra and H$\alpha$ and/or IR excess.
        $^c$Objects with KMOS spectra and low surface gravity and/or IR excess.
        $^d$Membership in \citet{meng17}.
    \end{center}
    \label{tab:tab_dataset_members}
\end{table}

In this section we perform the final assessment of the census of members of the core region of NGC 2244. So far, we found 107 candidate members from the literature (see Sect.~\ref{memb_high}) and 25 members and 10 possible members from the analysis of the KMOS spectra (see Sect.~\ref{analysis_youth}). Additionally, we find KMOS possible member \#77 to present infrared excess. This object only has IRAC1 and IRAC2 photometry, but it presents strong excess in both bands. We also find seven of the candidate members with VIMOS spectra that did not present accretion-related H$\alpha$ emission to present infrared excess. Two of the members that have KMOS spectra are also members in both \citet{muzic22} and \citet{meng17} catalogs (\#387, \#770), and another two members appear only in the \citet{meng17} catalog (\#18, \#152). KMOS member \#18 also has VIMOS spectrum giving the same SpT and extinction (within errors) as in the KMOS spectrum (see Sect.~\ref{analysis_kmos}). The VIMOS spectrum of object \#18 does not present accretion-related H$\alpha$ emission and the object does not present infrared excess either.

We consider a candidate member to be a bona-fide member if any of these conditions are met: (1) is a member in \citet{muzic22}, (2) has a VIMOS spectrum with accretion-related H$\alpha$ emission and/or infrared excess, (3) has a KMOS spectrum with low-gravity classification and/or infrared excess. In total, we obtain 117 bona-fide members. We also consider a candidate member to be a possible member if any of these conditions are met: (1) is a member in \citet{meng17} and did not meet any of the bona-fide conditions, (2) has a KMOS spectrum with a possible member classification based on the gravity-sensitive indices (see Sect.~\ref{analysis_youth}) and does not present infrared excess. In total, we obtain 22 possible members, one of which (\#208) meets both of the possible member conditions. In Table~\ref{tab:tab_dataset_members} we show how many of these bona-fide and possible members come from each member list.

\subsection{Derivation of stellar parameters}
\label{analysis_hrd}

\begin{figure}[hbt!]
    \centering
    \includegraphics[width=\textwidth/21*10]{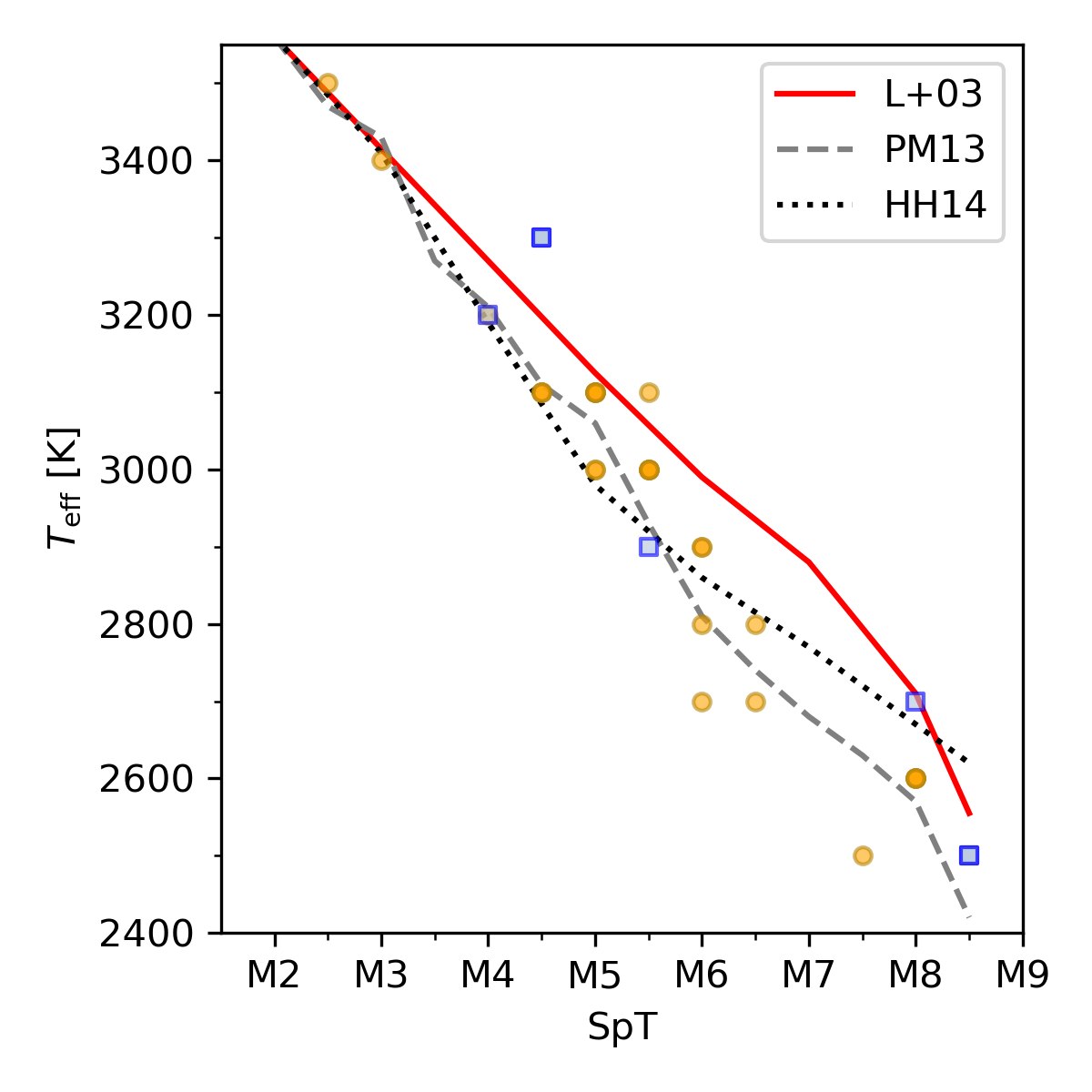}
    \caption{$T_{\mathrm{eff}}$ derived in Sect.~\ref{analysis_hrd} compared with the SpT derived in Sect.~\ref{analysis_templates} for the bona fide members (orange circles) and possible members (blue squares). The young  $T_{\mathrm{eff}}$ scale from \citet[L03, red solid line]{luhman03a}, the field $T_{\mathrm{eff}}$ scale from \citet[PM13, grey dashed line]{pecaut13}, and the young $T_{\mathrm{eff}}$ scale from \citet[HH14, black dotted line]{herczeg14} are also shown.}
    \label{fig:spt_teff}
\end{figure}

\begin{figure}[hbt!]
    \centering
    \includegraphics[width=\textwidth/21*10]{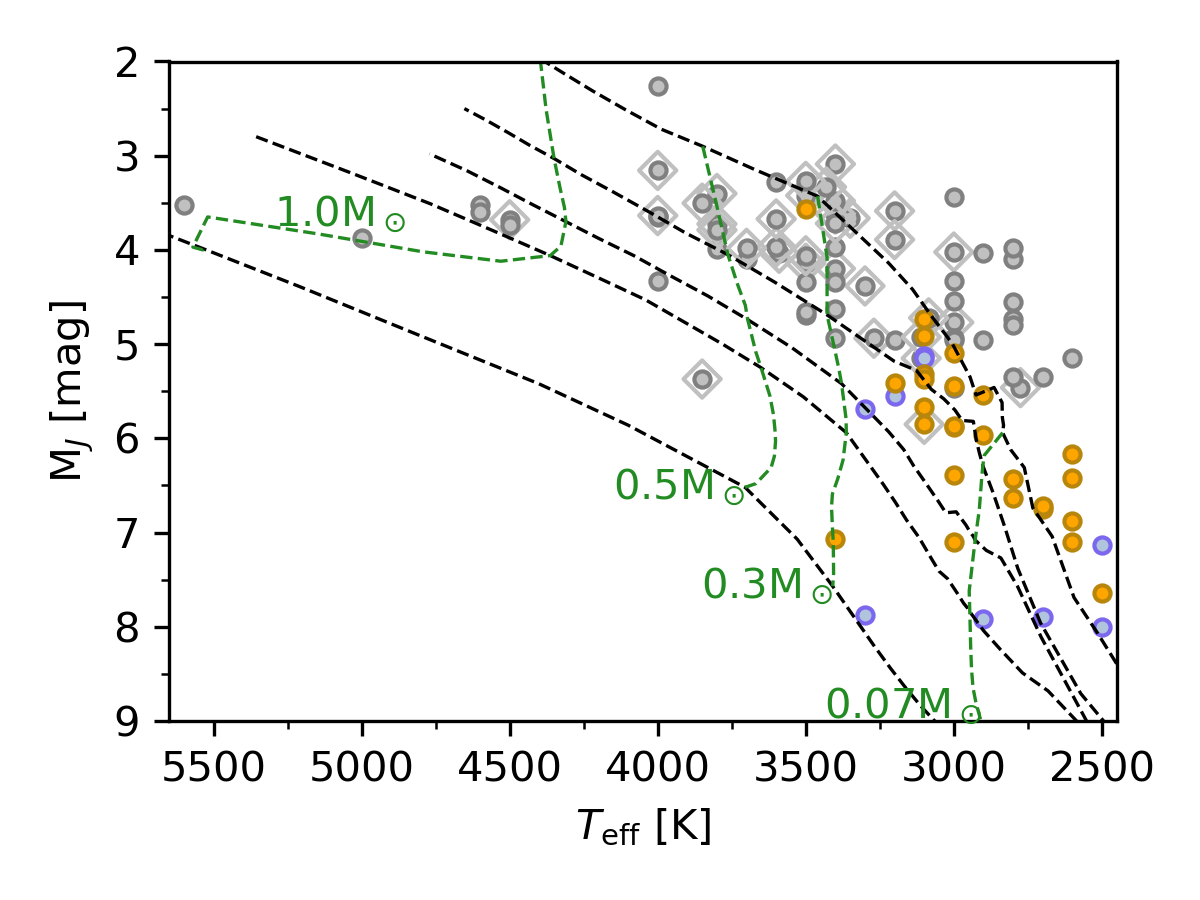}
    \caption{Hertzsprung-Russell diagram of all the members of NGC 2244. The color-coding is the same as in Fig.~\ref{fig:cmd}. The isochrones (black dashed lines, shown for 0.5, 2, 5, 10 and 100 Myr) and the lines of constant mass (green dashed lines) shown are from \citet{baraffe15}.}
    \label{fig:hrd_all}
\end{figure}

We derive the stellar parameters ($T_{\mathrm{eff}}$ and mass) of all the candidate members in the core region of NGC 2244 (see Sect.~\ref{analysis_excess}). For the KMOS members and possible members we derive $T_{\mathrm{eff}}$ together with log$g$ by direct comparison with the BT-Settl models \citep{allard12}. We de-redden all the spectra using the $A_{\mathrm{V}}$ derived in Sect.~\ref{analysis_templates}. We fix the metallicity at the solar value leaving only $T_{\mathrm{eff}}$ and log$g$ as free parameters. The model $T_{\mathrm{eff}}$ grid we use is 1500 -- 4200 K with a step of 100 K, encompassing the highest and lowest possible $T_{\mathrm{eff}}$ expected in our sample based on their SpT. We constrain the model log$g$ grid to 3.5 and 4 dex, since these are the values expected at the age of the cluster and the mass range of the candidate members based on the \citet{baraffe15} theoretical isochrones. The comparison is performed following the same steps of Sect.~\ref{analysis_templates}, but in this case the free parameters are $T_{\mathrm{eff}}$, log$g$ and the normalization wavelength. We adopt the $T_{\mathrm{eff}}$ from the best fit model. 

In Fig.~\ref{fig:spt_teff} we show the relationship between SpT and $T_{\mathrm{eff}}$ for our sample. We also show the young $T_{\mathrm{eff}}$--SpT scale from \citet{luhman03a} with a red solid line, the young scale from \citet{herczeg14} with a black dotted line and the field scale from \citet{pecaut13} \footnote{An updated version of this scale is available online: \url{https://www.pas.rochester.edu/~emamajek/EEM_dwarf_UBVIJHK_colors_Teff.txt}} with a grey dashed. We observe that while the three scales agree well with our results, the scale from \citet{pecaut13} agrees better with our results of the model fitting in the M6-M8 range (see Sect.~\ref{app_teff_scale} for further discussion). For the members from \citet{muzic22} we take the $T_{\mathrm{eff}}$ and $A_{\mathrm{V}}$ values given there coming from SED fitting. For the members and possible members with VIMOS spectroscopy, we use the $T_{\mathrm{eff}}$--SpT scale of \citet{pecaut13} to convert the SpT into $T_{\mathrm{eff}}$, and take the $A_{\mathrm{V}}$ given in \citet{muzic22} obtained from a template fitting process.

Using $A_{\mathrm{V}}$, a distance to the sources of 1500 pc and the photometry from the Flamingos-2 catalog we obtain the absolute $J$-band magnitude ($M\mathrm{_J}$). Combining $M\mathrm{_J}$ with $T_{\mathrm{eff}}$ we construct the Hertzsprung-Russell (HR) diagram of the entire dataset (see Fig.~\ref{fig:hrd_all}). We overplot the \citet{baraffe15} theoretical isochrones for ages between 0.5 and 100 Myr. We observe that most of the sources lie above the 2 Myr isochrone. Some recent studies suggest that the cluster may be younger than previously thought \citep[between 1 and 2 Myr;][]{wareing18,muzic22,roman23}, although these results are still in agreement with an age of 2 Myr. 

For each object, we associate the age and mass of the closest point of the isochrones interpolated to a finer grid of 1000 by 1000 points in age and mass respectively. The uncertainty of the derived parameters was estimated with a Monte Carlo (MC) simulation by varying $T_{\mathrm{eff}}$ and $M_{\mathrm{J}}$ by their respective uncertainties. For $T_{\mathrm{eff}}$ we assumed an uncertainty of 100 K (model grid step). For M$_J$ we include the photometric errors as well as an uncertainty of 0.25 mag in $A_{\mathrm{V}}$, added in quadrature. The distance uncertainty is not included in this calculation. The derived mass and age distributions for each object do not typically follow a normal distribution. Therefore, we save the derived distributions, which are later used to draw random samples for the derivation of the IMF. Objects \#205 (KMOS bona-fide member), \#368 (KMOS possible member) and \#119 \citep[member of][with VIMOS spectrum]{muzic22} are subluminous. Both \#205 and \#119 present infrared excess in the \textit{Spitzer} data, which could mean that their disk is seen edge-on \citep{crapsi08}.

There are 10 possible members coming from \citet{meng17} without VIMOS or KMOS spectroscopy for which we only have $JHK\mathrm{_S}$ photometry. We derive their mass and $A_{\mathrm{V}}$ by de-reddening the objects to the 2 Myr \citet{baraffe15} isochrone (shifted to the distance of NGC 2244, 1500 pc) in the $J-K_{\mathrm{S}}$ /$J$ diagram. In order to obtain the uncertainties of these measurements we perform a MC simulation by varying $J$ and $J-K_{\mathrm{S}}$ by their respective uncertainties. In Table~\ref{tab:tab_app_members} we provide the properties (coordinates, SpT, $A_{\mathrm{V}}$ and $T_{\mathrm{eff}}$) of all the members and candidate members in the core of NGC 2244 we have found. This table also includes the origin of their membership candidacy to the cluster, the infrared excess key derived in Sect.~\ref{analysis_excess} and the final membership assessment (see Sect.~\ref{analysis_census}).

\begin{table*}
    \caption{Properties of the observed KMOS objects with late SpT spectroscopic shape.}
    \begin{center}
        \begin{tabular}{r c c c c c c c c c c c c}
            \hline\hline
            ID & RA & DE & \textit{SpT} & $A_{\mathrm{V}}$ & \textit{SpT}$_{\mathrm{ind}}$ & $T_{\mathrm{eff}}$ & Mass & Low-g\tablefootmark{a} & Excess & Member \\ 
             & [º] & [º] & & [mag] &  & [K] & [$M_\odot$] &  &  &  \\ 
            \hline
            2 & 97.96073 & 4.95813 & M5 & 5.8 & M5.8 & 3100 & - & N & - & N\\ 
            5 & 97.96189 & 4.96056 & M5.5 & 6.6 & M5.5 & 3100 & 0.158$^{+0.023}_{-0.02}$ & Y & Y & Y\\ 
            6 & 97.96191 & 4.95555 & M6 & 3.0 & M6.6 & 2900 & 0.094$^{+0.016}_{-0.01}$ & Y & Y & Y\\ 
            16 & 97.96713 & 4.95692 & M4.5 & 3.4 & M5.7 & 3300 & - & N & - & N\\ 
            18 & 97.96784 & 4.97609 & M5 & 0.8 & M5.8 & 3100 & 0.153$^{+0.027}_{-0.028}$ & Y & N & Y\\ 
            29 & 97.97114 & 4.95991 & M5 & 4.6 & M4.7 & 3000 & - & N & - & N\\ 
            40 & 97.97569 & 4.94459 & M5.5 & 3.4 & M6.0 & 3000 & 0.11$^{+0.032}_{-0.021}$ & Y & - & Y\\ 
            48 & 97.97693 & 4.96129 & M4.5 & 2.6 & M6.4 & 3100 & 0.159$^{+0.024}_{-0.023}$ & ? & N & ?\\ 
            76 & 97.98241 & 4.92731 & M6.5 & 1.8 & M6.4 & 2700 & 0.039$^{+0.006}_{-0.004}$ & Y & Y & Y\\ 
            77 & 97.98241 & 4.95653 & M7.5 & 0.8 & M7.3 & 2500 & 0.019$^{+0.002}_{-0.002}$ & ?\tablefootmark{b} & Y & Y\\ 
            94 & 97.98552 & 4.92936 & M8 & 0.0 & M8.0 & 2600 & 0.046$^{+0.001}_{-0.002}$ & Y & N & Y\\ 
            107 & 97.98688 & 4.91758 & M6.5 & 2.8 & M5.0 & 2800 & 0.049$^{+0.011}_{-0.006}$ & Y & N & Y\\ 
            113 & 97.98774 & 4.93406 & M8 & 1.8 & M8.6 & 2600 & 0.029$^{+0.002}_{-0.001}$ & Y & Y & Y\\ 
            116 & 97.98921 & 4.90463 & M4.5 & 3.6 & M4.8 & 3100 & 0.157$^{+0.024}_{-0.024}$ & ? & N & ?\\ 
            124 & 97.99059 & 4.97549 & M5.5 & 1.4 & M5.9 & 3000 & 0.09$^{+0.023}_{-0.018}$ & Y & N & Y\\ 
            128 & 97.99173 & 4.90949 & M4 & 2.4 & M3.6 & 3200 & - & N & - & N\\ 
            152 & 97.99596 & 4.94247 & M8 & 0.2 & M7.1 & 2600 & 0.031$^{+0.003}_{-0.001}$ & Y\tablefootmark{d} & N & Y\\ 
            172 & 98.00105 & 4.92294 & M8.5 & 1.6 & M5.4 & 2500 & 0.016$^{+0.003}_{-0.001}$ & ? & N & ?\\ 
            205 & 98.00837 & 4.9626 & M3 & 0.4 & M3.7 & 3400 & 0.292$^{+0.066}_{-0.046}$ & Y & N & Y\\ 
            208 & 98.00915 & 4.96126 & M4 & 1.6 & M4.1 & 3200 & 0.2$^{+0.036}_{-0.028}$ & ?\tablefootmark{d} & N & ?\\ 
            209 & 98.00928 & 4.96238 & M5 & 2.0 & M6.3 & 3000 & 0.109$^{+0.031}_{-0.02}$ & Y & N & Y\\ 
            210 & 98.00967 & 4.91954 & M5.5 & 3.4 & M5.9 & 3000 & 0.097$^{+0.027}_{-0.019}$ & Y & N & Y\\ 
            211 & 98.01 & 4.92485 & M5.5 & 1.8 & M5.6 & 3000 & 0.13$^{+0.017}_{-0.009}$ & Y & N & Y\\ 
            230 & 98.01613 & 4.95503 & M5 & 3.0 & M5.8 & 3100 & 0.158$^{+0.028}_{-0.024}$ & Y & Y & Y\\ 
            234 & 98.01659 & 4.95427 & M5 & 9.6 & M5.7 & 3300 & - & N & - & N\\ 
            237 & 97.95597 & 4.9758 & M6 & 1.6 & M7.0 & 2900 & 0.078$^{+0.019}_{-0.008}$ & Y & N & Y\\ 
            285 & 97.97409 & 4.94141 & M6 & 2.6 & M5.5 & 2800 & 0.053$^{+0.008}_{-0.005}$ & Y & N & Y\\ 
            319 & 97.98101 & 4.98192 & M4.5 & 1.2 & M4.6 & 3300 & 0.25$^{+0.05}_{-0.038}$ & ? & N & ?\\ 
            338 & 97.98401 & 4.94803 & M4 & 3.2 & M3.6 & 3200 & 0.199$^{+0.032}_{-0.027}$ & Y & - & Y\\ 
            341 & 97.98467 & 4.97678 & M5 & 4.4 & M6.4 & 3100 & 0.16$^{+0.028}_{-0.03}$ & Y & N & Y\\ 
            354 & 97.98636 & 4.95939 & M5 & 1.6 & M5.7 & 3000 & 0.122$^{+0.028}_{-0.015}$ & Y & N & Y\\ 
            358 & 97.98677 & 4.97984 & M4.5 & 6.2 & M5.1 & 3200 & - & N & - & N\\ 
            368 & 97.98796 & 4.95934 & M4.5 & 1.6 & M7.6 & 3300 & 0.217$^{+0.045}_{-0.044}$ & ? & N & ?\\ 
            387 & 97.99153 & 4.92459 & M5 & 2.2 & M5.3 & 3100 & 0.159$^{+0.024}_{-0.014}$ & Y\tablefootmark{c}\tablefootmark{d} & Y & Y\\ 
            402 & 97.99456 & 4.91512 & M4.5 & 5.2 & M4.3 & 3100 & 0.159$^{+0.028}_{-0.027}$ & Y & Y & Y\\ 
            405 & 97.99476 & 4.95691 & M8.5 & 0.2 & M9.1 & 2500 & 0.027$^{+0.001}_{-0.001}$ & ?\tablefootmark{b} & N & ?\\ 
            412 & 97.99639 & 4.9322 & M8 & 1.2 & M6.8 & 2600 & 0.05$^{+0.0}_{-0.0}$ & Y & Y & Y\\ 
            415 & 97.99677 & 4.94423 & M6 & 1.2 & M7.0 & 2700 & 0.04$^{+0.004}_{-0.004}$ & Y & N & Y\\ 
            439 & 98.00067 & 4.94907 & M8 & 2.0 & M7.2 & 2700 & 0.029$^{+0.009}_{-0.006}$ & ?\tablefootmark{b} & N & ?\\ 
            567 & 97.98735 & 4.95598 & M5.5 & 2.8 & M7.2 & 2900 & 0.058$^{+0.015}_{-0.011}$ & ?\tablefootmark{b} & N & ?\\ 
            639 & 97.95719 & 4.97565 & M5 & 10.8 & M5.4 & 3300 & - & N & - & N\\ 
            770 & 97.97326 & 4.96323 & M2.5 & 2.6 & M3.0 & 3500 & 0.322$^{+0.035}_{-0.033}$ & Y\tablefootmark{c}\tablefootmark{d} & N & Y\\ 
            \hline 
        \end{tabular}
        \\$^a$Low surface gravity key defined in Sect.~\ref{analysis_youth} based on three gravity-sensitive spectral indices.
        $^b$Spectra where a low $S/N$ does not allow a confident derivation of gravity-sensitive spectral indices, they are classified as possible members.
        $^c$Classified as member in \citet{muzic22}.
        $^d$Classified as member in \citet{meng17}.
    \end{center}
    \label{tab:tab_results}
\end{table*}

\subsection{Completing the census}
\label{analysis_kmos_stat}

The KMOS member sample only represents a fraction of all the cluster members in its brightness range. In order to have a complete census we have to apply the obtained confirmation rate (number of confirmed members over number of observed objects) to the objects without spectroscopy in this brightness range. These ``new members'' together with the the bona-fide members will represent the underlying full population of the cluster. We define the KMOS brightness range to be $J$=16.5-19 mag. $J$=16.5 mag is the completeness limit of the \citet{muzic22} study, and $J$=19 mag roughly corresponds to 0.03 $M_\odot$ based on the distance, age and mean extinction of the cluster. However, below $J$=18.5 mag ($\sim$0.045 $M_\odot$) the quality of the KMOS spectra starts to worsen and we only find one bona-fide member (see right panel of Fig.~\ref{fig:cmd}). We may therefore have an incomplete census in the $J$=18.5-19 mag range. There are 143 objects in the Flamingos-2 catalog within the defined brightness range that were not observed with KMOS and are not members in any of the other catalogs.

The confirmation rate of the KMOS sample is found to be 36$\pm$2\%, which is in good agreement with the contamination rate of $\sim$60\% found in \citet{muzic19}.However, we observe that the confirmation rate changes significantly at different brightness levels, being $\approx$70\% at $J$$<$17.5 and $\approx$30\% at $J$$>$17.5 mag. We therefore follow a similar approach to that used in \citet{muzic19} for the statistical cluster membership (see their Sect. 4.2.2). The CMD is divided in several grid cells with a certain $\Delta$col and $\Delta$mag. In each cell grid we calculate the confirmation rate by integrating the contribution of each KMOS object to that cell and then obtaining the fraction of members. Each KMOS object is represented by a gaussian probability distribution, where the width is equal to its uncertainty. The measured in-cell confirmation rates are then applied to the objects in the Flamingos-2 catalog without spectra contained in each cell. Following \citet{muzic19} methodology, we vary both the size and position of the grid cells. We have tested a cell size of $\Delta$col=(0.3,0.4) mag and $\Delta$mag=(0.5,0.6) mag. Furthermore, for each combination of the color and magnitude cell sizes, we repeat the procedure with the grid shifted by $\pm$1/3 of the cell width in each dimension. Additionally, we perform the analysis on both the $J-K_{\mathrm{S}}$ /$J$ and $J-H$ /$J$ CMDs, resulting in 72 different grid cell configurations. Since this membership extrapolation method is statistical, we obtain 1000 different member lists by bootstrapping the 72 grid cell configurations.

We derive the mass and $A_{\mathrm{V}}$ of all the sources in the KMOS brightness range without spectroscopy. We use the same methodology used in Sect.~\ref{analysis_hrd} for the \citet{meng17} possible members de-reddening to the 2 Myr \citet{baraffe15} isochrone in the $J-K_{\mathrm{S}}$ /$J$ diagram. The mass completeness of the combined census is given by the magnitude limit imposed in the membership statistical extrapolation, which is roughly equivalent to 0.03 $M_\odot$. However, as mentioned previously, this limit may be overestimated. Therefore, we set the completeness limit of the spectroscopic observations to $J$=18.5 mag, i.e. 0.045 $M_\odot$. While it is not straightforward to assess the completeness limit of a spectroscopic follow up, we opted for setting this rather conservative limit to avoid artifacts due to incompleteness at lower masses.

\section{Results}
\label{imf}
\subsection{Initial mass function}

\begin{figure}[hbt!]
    \centering
    \includegraphics[width=\textwidth/21*10]{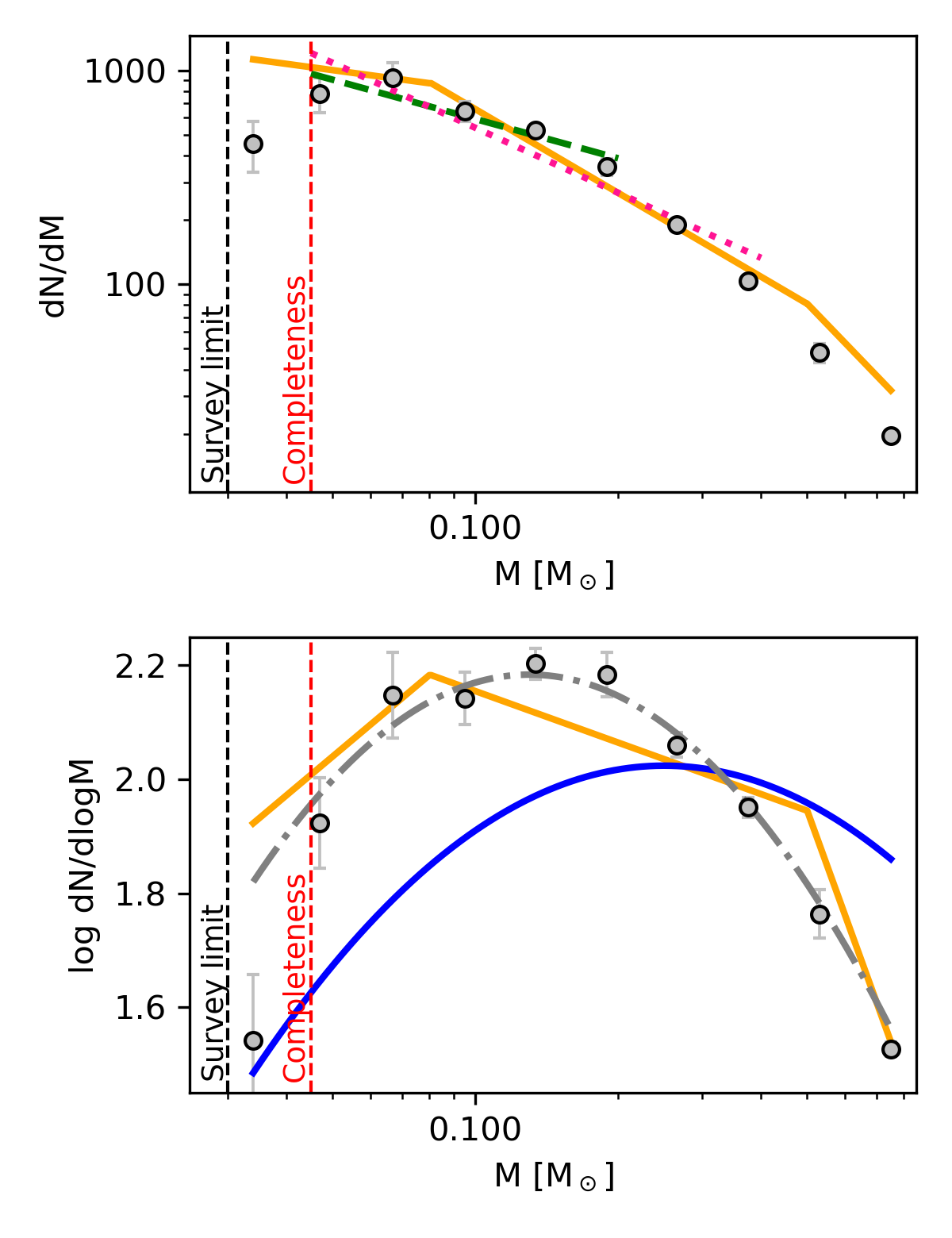}
    \caption{IMF of the central region of NGC 2244 considering both KMOS and \citet{meng17} possible members with P=0.5 (see text), represented in equal-size logarithmic bins with size $\Delta$log($m/M_\odot$)=0.15. Two different IMF representations are shown: $dN/dm$ (top panel) and $dN/dlog(m)$ (bottom panel). The green dashed and the pink dotted lines in the top panel represent the power-law fits performed between 0.045-0.2 $M_\odot$ and 0.045-0.4 $M_\odot$ respectively. The orange solid line (both panels) represents the Kroupa segmented power-law mass function \citep{kroupa01}, the blue solid line (bottom panel) shows the Chabrier mass function \citep{chabrier05} and the grey dash-dotted line (bottom panel) represents a log-normal function with a characteristic mass ($m\mathrm{_c}$) of 0.13 $M_\odot$ and standard deviation ($\sigma$) of 0.45. The three representations of the IMF are normalized to match the total number of objects in the cluster. The black vertical dashed line at 0.03 $M_\odot$, and the red line at 0.045 $M_\odot$ represent the KMOS survey limit and the true mass completeness of our sample respectively.}
    \label{fig:imf}
\end{figure}

\begin{table}
    \caption{IMF $\alpha$ slopes for the 9 treatments of the possible members calculated over two different mass ranges: 0.045-0.2 $M_\odot$ and 0.045-0.4 $M_\odot$.}
    \begin{center}
        \begin{tabular}{l c c c c}
            \hline\hline
            \multirow{2}{*}{KMOS\tablefootmark{a}} & \multirow{2}{*}{M17\tablefootmark{a}} & \multicolumn{2}{c}{IMF $\alpha$ slope} \\ 
             &  & 0.045--0.2 M$_\odot$ & 0.045--0.4 M$_\odot$ \\ 
            \hline
            all & all & 0.68 $\pm$ 0.11 & 1.05 $\pm$ 0.15 \\ 
            random & all & 0.68 $\pm$ 0.11 & 1.04 $\pm$ 0.15 \\ 
            none & all & 0.7 $\pm$ 0.11 & 0.99 $\pm$ 0.13 \\ 
            all & random & 0.69 $\pm$ 0.12 & 1.08 $\pm$ 0.16 \\ 
            random & random & 0.7 $\pm$ 0.12 & 1.08 $\pm$ 0.15 \\ 
            none & random & 0.71 $\pm$ 0.11 & 1.03 $\pm$ 0.13 \\ 
            all & none & 0.68 $\pm$ 0.13 & 1.11 $\pm$ 0.17 \\ 
            random & none & 0.7 $\pm$ 0.13 & 1.11 $\pm$ 0.16 \\ 
            none & none & 0.73 $\pm$ 0.13 & 1.06 $\pm$ 0.14 \\ 
            \hline 
            \multicolumn{2}{c}{Mean} & 0.7 $\pm$ 0.12 & 1.06 $\pm$ 0.16 \\ 
            \hline 
        \end{tabular}
        \\$^a$Treatment of the possible members: ``all'' means P=1, ``random'' means P=0.5, and ``none'' means P=0.
    \end{center}
    \label{tab:tab_imf_slopes}
\end{table}

The IMF is obtained using the methodology presented in \citet{muzic22}. The mass of each member is represented by a distribution derived from the different MC simulations. Each of these distributions is smoothed by a Gaussian kernel density estimator (KDE), and used to extract a random sample of N$_1$ values per object, constructing thus N$_1$ realisations of the mass distributions for the entire sample. For each of these realisations, we perform N$_2$ bootstraps, resulting in N$_1 \times$N$_2$ mass distributions. Each distribution is then binned onto the same grid, and the final value and its uncertainty for each bin is calculated as mean and standard deviation of N$_1 \times$N$_2$ values. We set N$_1$=N$_2$=100. The masses are binned to logarithmic bin sizes of $\Delta$log($m/M_\odot$) = 0.2 for masses between 0.03 and 1 $M_\odot$. The final IMF is derived from the mean and standard deviation of the IMF obtained using the 1000 different member lists. Furthermore, we repeat this procedure for different treatments of the possible members. We weighted the possible members with three different probabilities, that is P = 0,1, and 0.5. We considered these three scenarios separately for the KMOS and the \citet{meng17} possible members, exploring 9 different scenarios for the treatment of the possible members. 

In order to minimize the dependence of the IMF slopes derived on the choice of the bin size and location, we repeat the IMF calculation for two additional bin sizes, $\Delta$log($m/M_\odot$) = 0.15 and 0.25. We also repeat the calculation with the same bin size as before ($\Delta$log($m/M_\odot$) = 0.2) but shift the location of the bins by half the bin size. In total we derive 36 different IMFs. In Fig.~\ref{fig:imf} we show the IMF obtained when including the KMOS and \citet{meng17} possible members with 0.5 probability of being members in each member list with bin sizes of $\Delta$log($m/M_\odot$)=0.15. In the top panel of Fig.~\ref{fig:imf} we observe that the mass bin located below the mass completeness limit at 0.045 $M_\odot$ (red dashed line), is well below the power law fits we performed and the Kroupa mass function. This confirms the unreliable nature of the membership methodology below this limit. The treatment of the KMOS possible members affects all mass bins below 0.3 $M_\odot$, specially the last one, where possible members dominate. The treatment of the \citet{meng17} possible members affect the mass bins between 0.2 and 0.8 $M_\odot$. 

In the bottom panel of Fig.~\ref{fig:imf} we plot the IMF in the log-normal form ($dN/dlogM$) together with the Kroupa \citep[orange line]{kroupa01} and Chabrier \citep[blue line]{chabrier05} mass functions. Both mass functions are normalized to have the same number of objects as the IMF derived for NGC 2244. We observe that the peak mass of the IMF is lower and the spread smaller than predicted by the Chabrier mass function, whereas the Kroupa mass function reproduces the overall shape of the IMF. We find the log-normal form of the IMF to be well-reproduced by a single log-normal function (grey dash-dotted line) down to the completeness limit of the observations. The shown log-normal function has a characteristic mass (peak of the mass distribution) of 0.13 $M_\odot$ and a standard deviation of 0.45. This characteristic mass is lower than that found for NGC 2244 in \citet[$\sim$0.23-0.28~$M_\odot$]{damian21}. However, the completeness of the observations of NGC 2244 in \citet{damian21} is $\sim$0.07~$M_\odot$. It is also lower than that found in nearby SFRs \citep[see][and references therein]{suarez19}. A possible reason for this difference is that we have not consider objects with masses larger than 1 $M_\odot$, which could potentially shift the characteristic mass of the IMF of NGC 2244 towards larger values.

We fit the 36 IMFs with power laws ($dN/dM \propto M^{-\alpha}$) at three different mass ranges: 0.045 -- 0.2 $M_\odot$, 0.045 -- 0.4 $M_\odot$, 0.045 -- 1 $M_\odot$. The fit performed up to 1 $M_\odot$ produces an $\alpha$ that is larger than any previously reported $\alpha$ value (1.41$\pm$0.15). However, we observe that this fit is well above the lowest mass bins, which means that the $\alpha$ slopes derived from this fit are overestimated. We do not discuss these results further since they are not representative of the low-mass population of the cluster.  The other two fits are able to reproduce the shape of the IMF in their mass ranges and are shown in the top panel of Fig.~\ref{fig:imf} as green dashed and pink dotted lines respectively. In Table~\ref{tab:tab_imf_slopes} we present the $\alpha$ slopes obtained in the two mass ranges for the nine treatments of the possible members. Here, the results for the different mass bins have been combined together. We also provide the mean and standard deviation of the $\alpha$ for the two mass ranges explored. 

The $\alpha$ slope we found below 0.2 $M_\odot$ ($\alpha$=0.7$\pm$0.12) agrees well with the previous result found for NGC 2244 in \citet{muzic19} for the 0.02-0.1 $M_\odot$ mass range ($\alpha$=0.75$\pm$0.13). While the result we obtain below 0.4 $M_\odot$ ($\alpha$=1.06$\pm$0.16) agrees with the results for the 0.02-0.4 $M_\odot$ mass range in \citet[$\alpha$=1.03$\pm$0.03]{muzic19}. The 0.045-0.4 $M_\odot$ slope is steeper than the slope found in \citet{muzic17} for RCW 38 ($\alpha$=0.7$\pm$0.1 for the mass range 0.02–0.5 $M_\odot$). In nearby SFRs at distances of up to 400 pc, the $\alpha$ slope typically ranges between 0.5-1 \citep[][and references therein]{muzic17,gennaro20,suarez19}. The 0.045-0.4 $M_\odot$ slopes we derived are on the high-end of production of low-mass stars and BDs, but they are still in agreement taking into account the uncertainties.

An environmental property that may affect the slope of the IMF is metallicity \citep{kroupa02,yan20,jiadong23}. Several studies have found a bottom-light IMF in regions with very low metallicity \citep{reyle01,geha13,jiadong23}. We cross-matched our members with measurements coming from the LAMOST \citep{luo15} and Gaia-ESO \citep{randich22} surveys. We obtain a metalliticy of [Fe/H]=-0.08$\pm$0.11 and [Fe/H]=-0.07$\pm$0.2 for LAMOST and Gaia-ESO respectively. These results are in agreement with the recent results derived for the entire Rosette nebula from APOGEE of [Fe/H]=-0.1$\pm$0.04 \citep{roman23} and with a solar metallicity.

\begin{table}
    \caption{Star-to-BD ratios for the 9 treatments of the possible members.}
    \begin{center}
        \begin{tabular}{l c c c c}
            \hline\hline
            \multirow{2}{*}{KMOS} & \multirow{2}{*}{M17} & \multicolumn{2}{c}{Star-to-BD ratio} \\ 
             &  & 0.03--1 M$_\odot$\tablefootmark{a} & 0.03--1 M$_\odot$\tablefootmark{b} \\ 
            \hline
            all & all & 2.9 $\pm$ 0.5 & 2.2 $\pm$ 0.3 \\ 
            random & all & 2.8 $\pm$ 0.5 & 2.2 $\pm$ 0.3 \\ 
            none & all & 2.9 $\pm$ 0.5 & 2.4 $\pm$ 0.4 \\ 
            all & random & 2.8 $\pm$ 0.5 & 2.2 $\pm$ 0.3 \\ 
            random & random & 2.8 $\pm$ 0.5 & 2.2 $\pm$ 0.3 \\ 
            none & random & 2.8 $\pm$ 0.5 & 2.3 $\pm$ 0.3 \\ 
            all & none & 2.8 $\pm$ 0.5 & 2.1 $\pm$ 0.3 \\ 
            random & none & 2.7 $\pm$ 0.5 & 2.1 $\pm$ 0.3 \\ 
            none & none & 2.8 $\pm$ 0.5 & 2.2 $\pm$ 0.3 \\ 
            \hline 
            \multicolumn{2}{c}{Mean} & 2.8 $\pm$ 0.5 & 2.2 $\pm$ 0.3 \\ 
            \hline 
        \end{tabular}
        \\$^a$Corrected for completeness down to 0.03 M$_\odot$ using the $<$0.2M$_\odot$ IMF fit.
        $^b$Corrected for completeness down to 0.03 M$_\odot$ using the $<$0.4M$_\odot$ IMF fit.
    \end{center}
    \label{tab:tab_stbds}
\end{table}

\subsection{Star/BD ratio}

We estimate the ratio between the number of stars and BDs (star-to-BD number ratio) using the same methodology as we did in deriving the IMF: for each of the 1000 member lists we extract N$_1$ mass values from the gaussian KDE of the mass distribution, then for each of the N$_1$ realizations, we perform N$_2$ bootstraps. We derive the star-to-BD number ratio and its uncertainty as the mean and standard deviation of the 1000 member lists. We place the star-BD boundary at 0.075 $M_\odot$ \citep{chabrier22}. We derive the ratio between 0.045 $M_\odot$ up to 1 $M_\odot$, and we do it for the nine different treatments of the possible members studied. We calculate the star-to-BD number ratio in this mass range because it is the range used in most measurements of the ratio in the literature. We find a mean star-to-BD number ratio of 3.6$\pm$0.9. However, most studies in the literature calculate the star-to-BD number ratio down to 0.03 $M_\odot$. Therefore, we derived the star-to-BD number ratio down to 0.045 $M_\odot$, and in the 0.03-0.045 $M_\odot$ mass range, we added the contribution from the IMF power law fits (given in Table~\ref{tab:tab_imf_slopes}). We find a mean star-to-BD number ratio of 2.8$\pm$0.5 and 2.2$\pm$0.3 when correcting for completeness using the IMF fit $<$0.2 $M_\odot$ and $<$0.4 $M_\odot$, respectively. We present these results in Table~\ref{tab:tab_stbds}. These values are in agreement with the mean value found in \citet[][3$\pm$0.3]{muzic19} for the core of NGC 2244. They are also in agreement to the star-to-BD number found in RCW 38 \citep[3.0$\pm$1;][]{muzic17} and the values typically found in nearby star forming regions \citep[2-5, e.g.][]{scholz13,muzic15,andersen08}, but at the same time, they are on the high-end of BD production. 

\begin{figure}[hbt!]
    \centering
    \includegraphics[width=\textwidth/21*10]{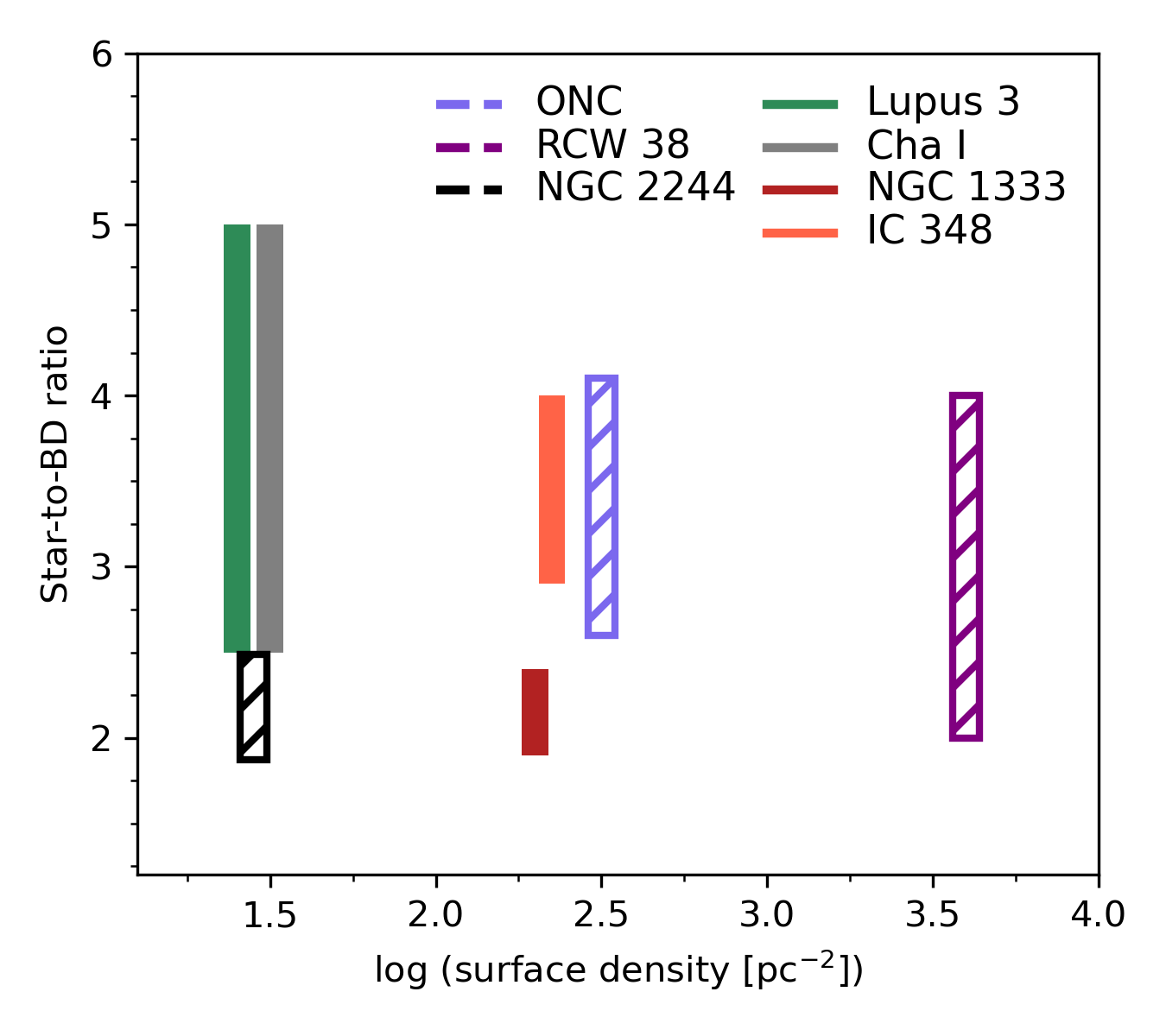}
    \caption{Dependence of the star-to-BD number ratio on cluster surface density. Different regions are represented by polygons of different colors. The height of each polygon represents either the $\pm$1$\sigma$ range around the mean value or the range in star-to-BD number ratio if given in this form. The width of the polygons has no physical meaning. The filled polygons represent the regions with few or no massive stars, while the dashed ones mark the regions with substantial OB star population. The result shown for NGC 2244 is the one calculated correcting for completeness with the IMF power law fit performed between 0.045 and 0.4 $M_\odot$.}
    \label{fig:star_to_bd}
\end{figure}

We take the stellar densities derived in \citet{muzic19} for several star forming regions and their star-to-BD number ratios from the literature and compare them with the values derived in this work. In Fig.~\ref{fig:star_to_bd} we show the values of the star-to-BD number ratio for different star forming regions as a function of the stellar surface density. Regions with high number of OB stars are shown with a hatched pattern. We represent the star-to-BD number ratio obtained when correcting for completeness using the $<$0.4 $M_\odot$ IMF power law fit with a black hatched polygon. The star-to-BD number ratio does not seem to depend on either the stellar density or the presence of OB stars. However, we observe that NGC 2244 hosts one of the (if not the) richest population of BDs ever found in a SFR.

\subsection{Distance to OB stars}
\label{results_distance}

\begin{figure}[hbt!]
    \centering
    \includegraphics[width=\textwidth/21*10]{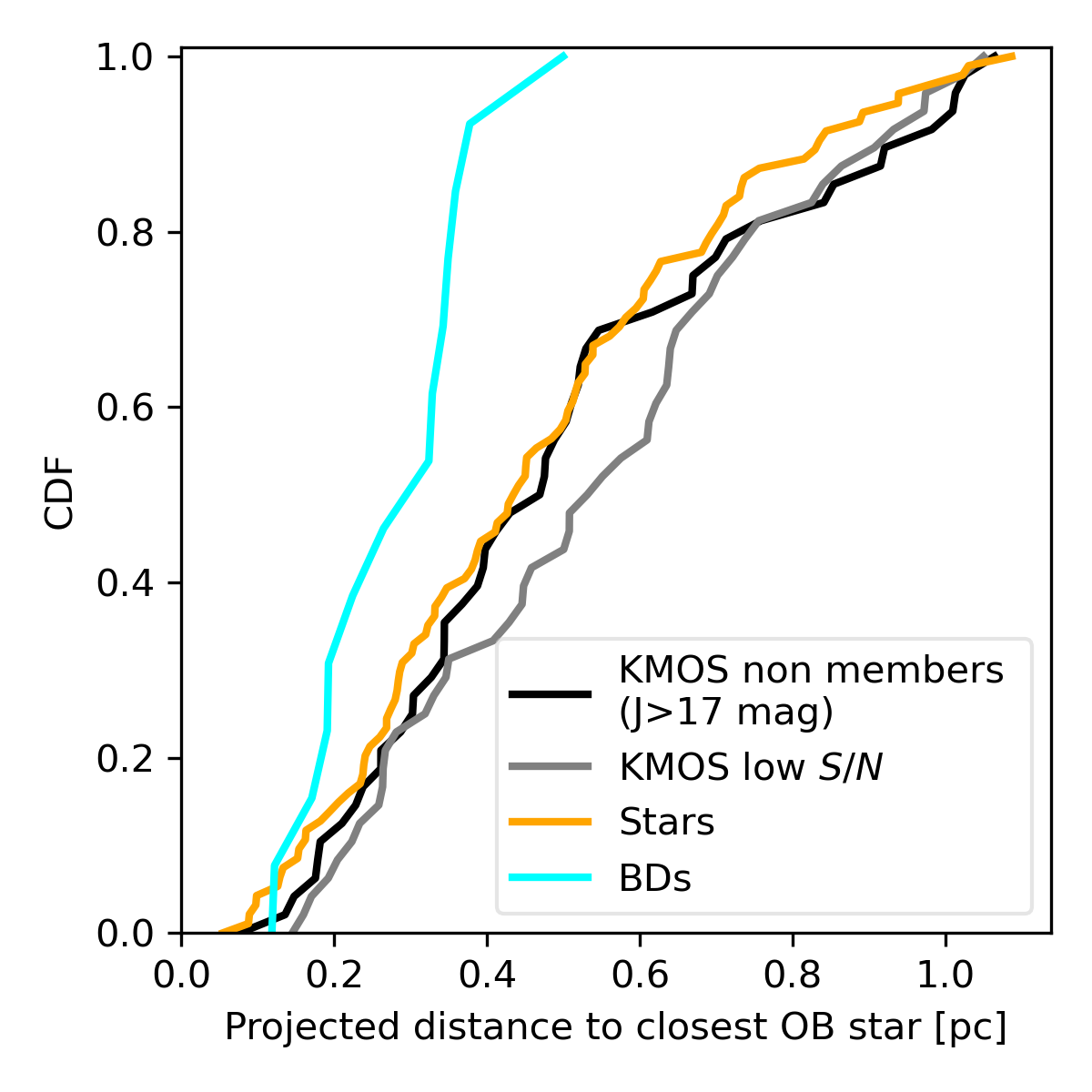}
    \caption{Cumulative distribution function of the projected distance to the closest OB star of the stars (orange line) and BDs (cyan line) members of the core of NGC 2244. We also show the distribution of non members with $J>$17 mag (black line) as well as the KMOS sources with very low $S/N$ that were not considered in any part of the analysis (grey line).}
    \label{fig:ob_distance}
\end{figure}

\begin{figure}[hbt!]
    \centering
    \includegraphics[width=\textwidth/21*10]{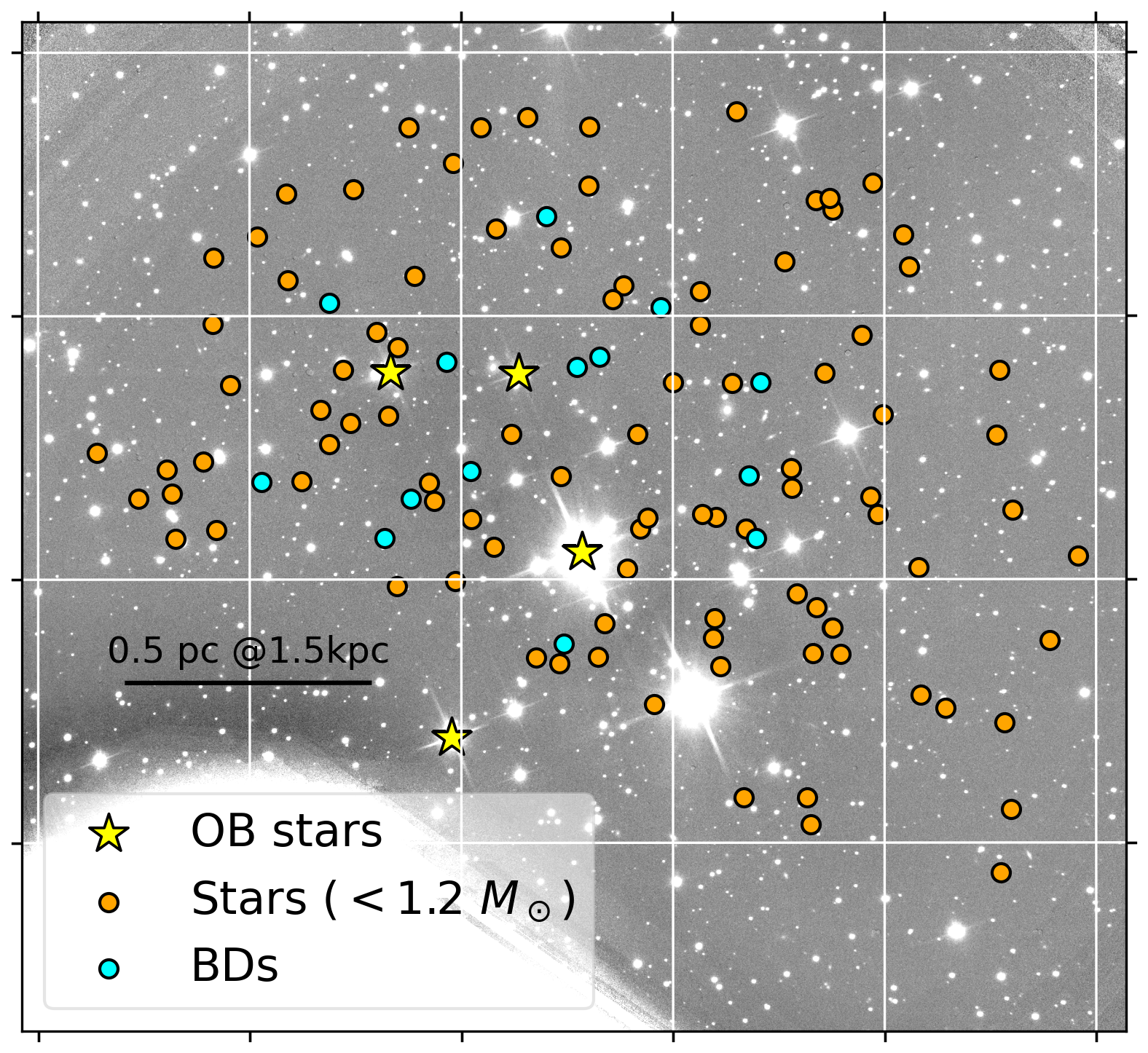}
    \caption{Sky distribution of all the members of the core region of NGC 2244. Orange circles represent stars  ($<$1.2 $M_\odot$), blue circles represent BDs and yellow stars represent the OB stars in the region coming from \citet{preibisch08}. On the background we show the $K_{\mathrm{S}}$-band image of the center of NGC 2244 taken with Flamingos-2/Gemini-S \citep{muzic19}.}
    \label{fig:ob_sky}
\end{figure}

Here we explore the distribution of projected distances between the OB stars and the stellar and substellar population of NGC 2244. We get the OB stars of the cluster from \citet{sfh_rosette} and assume that all members and OB stars are located 1500 pc from us. We then calculate the minimum projected distance of all the members to the OB stars in the cluster. In Fig.~\ref{fig:ob_distance} we show the projected distance to OB stars cumulative distribution function (CDF) of stars (orange solid line) and BDs (cyan solid line). We observe that BDs are typically closer to OB stars than stars. We performed an Anderson–Darling test to study the null hypothesis that the two samples are drawn from the same underlying population, obtaining a probability of 0.7\% that they are drawn from the same distribution, which confirms that the two distributions are different. This result holds when only the bona-fide members are used in the calculation (i.e. not including the possible members), and also when calculating the distance to the O-type stars (i.e. leaving the B-type stars out). In order to test that this result is not an observational bias in the sense that most of the faint sources with good KMOS spectra were by chance in the fields located closer to OB stars, in Fig.~\ref{fig:ob_distance} we also show the projected distance CDF of the KMOS faint non members ($J>$17 mag) and the observed KMOS sources that did not have enough $S/N$ for the analysis. We observe that these sources present a CDF much more similar to that of stars. The Anderson-Darling test confirms that these distributions are indeed very similar, finding a probability of 29.2\% and 17.2\% that stars and the faint non-members or the sources with insufficient $S/N$ are drawn from the same distribution, respectively.

In Fig.~\ref{fig:ob_sky} we show the spatial distribution of all the members of the core region of the cluster and that of the OB stars coming from \citet{sfh_rosette}. We observe that the stars are located all over the studied area, whereas the BDs seem to be preferentially located in the central region of the field. The overall asymmetry of both stellar and substellar members seen in Fig.~\ref{fig:ob_sky} is a consequence of the footprint of the Flamingos-2 observations \citep{muzic19}. This asymmetry should not affect our results, since it should have the same effect in the distribution of both stars and BDs. Mass segregation could also be affecting the spatial distribution of the cluster members. Several authors have explored mass segregation in NGC 2244 and found contradictory results \citep{chen07,wang08,muzic22}. \citet{muzic22} found that some degree of mass segregation may be present in NGC 2244, but it does not seem to be very pronounced.

\subsection{Disk fraction}
\label{results_df}

Due to the particular environment of the cluster (high number of OB stars while keeping nearby SFR-like stellar density), we can probe whether OB stars have had any impact in the evolution of protoplanetary disks in this cluster. We explored the fraction of members of NGC 2244 that present infrared excess (i.e. that have disks) evaluated in Sect.~\ref{analysis_excess}. We compute the disk fraction as the fraction between the number of sources that present disk excess and the total number of sources with IRAC data in at least two channels (analogous to: Class IIs over Class IIs and Class IIIs). We find a total disk fraction of 39$\pm$9\% (error coming from binomial probability distribution with Clopper–Pearson method). We find the disk fraction of stars (40$\pm$9\%) to be higher than that of BDs (31$\pm$22\%), although the errors in the disk fractions are very large to confirm any real difference. We observe no difference in the distribution of the projected distance to the closest OB star between the disk-bearing and diskless sources, and neither do we find any difference when separating the sample in stars and BDs. We find the same result when limiting the analysis to the O-type stars.

The \textit{Spitzer} data are available in at least two bands for most sources ($>$96\% of all the candidate members). However, the completeness of the IRAC3 and IRAC4 bands decreases significantly below $J$=17 mag. The 50\% completeness is located at $J$=18.5 mag in IRAC3 ($\sim$0.045~$M_\odot$), and at $J$=17 mag in IRAC4 ($\sim$0.1~$M_\odot$). This implies that we may be missing some BDs that only present disk excess in IRAC3 or IRAC4 and may account for the difference in the disk fraction we have found between stars and BDs. The same issue may be affecting the derived total disk fraction. Given the infrared excess detection probabilities provided in Table B.1, the possibility to not detect the infrared excess of a source with ``full'' disk but no detection at longer wavelengths than IRAC2 is below 30\%. Based on this, and taking into account the sources that only present \textit{Spitzer} detections in the IRAC1 and IRAC2 bands, the disk fraction may be higher by 1.5\%, which is well inside the uncertainty of the disk fraction derived. We also do not include in this analysis redder mid-infrared observations (i.e. 24 $\mu m$), so we are most likely missing some evolved disks that start dominating the emission at longer wavelengths (see Fig.~\ref{fig:app_ir_colors_cdf}). However, most of the disk fractions available in the literature are also suffering from the same caveat, either because they do not include young objects with evolved disks in the derivation of disk fractions or because they do not include any 24 $\mu m$ data in their analysis neither.

If we compare the disk fraction of NGC 2244 with that of nearby SFRs (see e.g. Fig. 5 of \citealt{michel21}), we observe that most nearby SFRs of ages $\sim$2 Myr have disk fractions $\geq$50\%, although our result in NGC 2244 is still in agreement within 1$\sigma$ with regions like Taurus or IC 348. \citet{pfalzner22} found that protoplanetary disks have median disk lifetimes of 5-10 Myr, and they proposed that the main reason behind the low disk mean lifetimes found until recently \citep[1-3 Myr,][]{cieza07,fedele10}, is that they were found targeting clusters located further away having therefore an over-representation of higher-mass stars that are known to disperse their disks earlier \citep{williams11,ribas15}. However, in this work we restricted the analysis to objects with SpT later than K0 and even extend into the BD regime. 

One possible explanation for the low prevalence of disks found in NGC 2244 is the presence of OB stars which would help disperse the protoplanetary disks faster with their strong UV radiation \citep[see review by][]{winter22}. The population of young stellar objects with disks in NGC 2244 has been previously studied using \textit{Spitzer} data \citep{balog07,poulton08}. \citet{balog07} found NGC 2244 to have a disk fraction of 44.5\%, while they found a lower disk fraction of 27\% when considering only the stars closer to O-type stars ($<$0.5 pc). This result hints that OB stars may be affecting the evolution of protoplanetary disks closest to them. Our result agrees within errors with both results found in \citet{balog07}, but we did not find a lower disk fraction for the members closest to the OB stars.

\section{Summary and conclusions}
\label{summary}

In this work we presented an analysis of the low-mass population of the core of NGC 2244 (central 2.4~pc$^2$) using new, deep NIR KMOS/VLT spectroscopy of faint candidate members. We also presented a new reduction of \textit{Spitzer} data in the region that goes into the BD regime. In the stellar regime the members come from a catalog built using optical to NIR photometry and \textit{Gaia} proper motions \citep{muzic22}, VIMOS/VLT optical spectroscopy and a catalog built using infrared excess and X-rays \citep{meng17}. We performed the analysis of the KMOS spectra, including the derivation of the stellar parameters (SpT, $A_{\mathrm{V}}$) and the assessment of their membership to NGC 2244 using gravity-sensitive spectral indices. We use the \textit{Spitzer} photometry to evaluate the infrared excess of all the candidate members of the cluster. From the KMOS spectra we find 26 new low-mass members and 9 possible members ($<$0.4~$M_\odot$) of NGC 2244, including the first spectroscopically confirmed free-floating BDs beyond 1 kpc, to the best of our knowledge.

We derived the IMF of the core region of NGC 2244, and find it to be well represented by a single power law between 0.045-0.2~$M_\odot$ and 0.045-0.4~$M_\odot$. The slope of the IMF is heavily dependent on the fitted mass range, with typical values $\alpha$=0.7-1.1. These $\alpha$ slopes are at the high-end, but still in agreement with those found in nearby SFRs ($\alpha$=0.5-1). We find the IMF to also be in agreement with the results found in \citet{muzic19} for the same region using only photometry. While our census is complete down to $\sim$0.045~$M_\odot$, most studies in the literature calculate the star-to-BD number ratio down to 0.03~$M_\odot$. In order to make a proper comparison with other works, we correct our census using the IMF power law fits, and  obtain a star-to-BD number ratio between 0.03 and 1~$M_\odot$ of 2.8$\pm$0.5 and 2.2$\pm$0.3 by correcting for completeness using the IMF fits below $<$0.2~$M_\odot$ and $<$0.4~$M_\odot$, respectively. These results are in agreement with previous results in nearby SFRs (2-5), although on the low end of typical values, confirming the rich population of BDs in the core of the cluster. As discussed in \citet{muzic19}, the BD population of massive young cluster NGC 2244 seems to agree very well with the results coming from nearby SFRs. If there are any variations in the low-mass mass function, they must be more subtle than the error bars allow us to discern.

We found that BDs in NGC 2244 are preferentially closer to the OB stars than stars. \citet{whitworth04} proposed that BDs can form due to the photoevaporation of the outer layers of the forming protostar by a nearby OB star. This mechanism is considered inefficient (a massive core is necessary to form a single BD) and cannot be a dominant process in BD formation (many BDs are observed in nearby SFRs where there are no OB stars), but it could still influence the production of BDs in centers of massive clusters rich in OB stars like NGC 2244. BDs formed via this mechanism will initially be in the vicinity of OB stars, so our finding seems to suggest that OB stars have stimulated the formation of BDs in the central region of NGC 2244. However, in this scenario at least a fraction of the BDs we have found, must have formed following the typical BD formation process in nearby SFRs. These BDs should have been distributed randomly along the cluster, and not preferentially closer to the OB stars. Given the current velocity dispersion of the cluster \citep[$\sim$1.4 km/s,][]{muzic22}, a typical NGC 2244 member will take $\sim$1.1 Myr to cross the core of the cluster (i.e. the region studied in this work, 2.4 pc$^2$). This implies that the initial spatial configuration of the members of NGC 2244 can be significantly different to the one we have observed. If there are BDs in NGC 2244 that have formed thanks to the effect of nearby OB stars, we would expect the BD population of NGC 2244 to be richer than that found in nearby SFRs where there are no OB stars. However, the range of sizes of the known BD populations is very wide, and the increase in BD formation due to OB stars could still keep NGC 2244 in agreement with previous results. A possible explanation for the lack of BDs in some regions of the cluster is that these regions present higher extinction that prevented us from finding BDs. However, the spatial distribution of the faint objects ($J$~$>$17 mag) with KMOS spectra that we find to be contaminants in Sect.~\ref{analysis_visual} is similar to that of stars. We also do not find evidence of systematic higher extinctions in the regions where we do not find BDs. Other important caveats of this result are that we are missing the line-of-sight positions of the cluster members and that the total number of BDs in this study is relatively low.

Another scenario proposed where the formation of BDs is facilitated by the presence of OB stars is by the gravitational collapse of globulettes. Globulettes are very small clumps of gas found in HII regions that, if they undergo gravitational collapse, would give birth to objects below the hydrogen-burning limit. These globulettes can be seen as a scaled-down version of the well-known proplyds \citep{odell93,johnstone98}, although they do not always have an object inside \citep{makela14,grenman14}. It has been proposed that these objects may be the detached or fragmented remnants of pillars located on the edges of HII regions \citep{gahm07,haworth15} and a rich population of globulettes has been found in the Rosette nebula \citep{gahm07,gahm13,makela14}. However, \citet{haworth15} found that globulettes alone are in principle stable against gravitational collapse, until they collide with the HII region shell. In some occasions, this impact may trigger the gravitational collapse of the globulette, therefore forming a BD. Following this reasoning, the collapse of these objects would occur far from the OB stars and would not be a possible scenario to explain our results. Another issue with this scenario is that the mass distribution of globulettes peaks below the deuterium-burning limit \citep[$\sim$13$M\mathrm{_{Jup}}$,][]{gahm07,gahm13}. If these objects undergo gravitational collapse, most of the objects formed will be on the planetary-mass domain. 

Here we discuss other proposed formation scenarios of BDs and whether they can explain the observed spatial distribution of BDs in NGC 2244. \citet{reipurth01} proposed that the dynamic interaction between protostars or fragments within a massive core, may result in the ejection of the smallest fragments. If the ejected bodies have masses below the hydrogen-burning limit they may become a BD. The ejection velocity of these objects would be of the order of 3 km/s, travelling a distance of 6 pc in 2 Myr (age of NGC 2244), which is inconsistent with our results. \citet{bate12} found that the velocity dispersion of the ejected clumps should also be larger than 1 km/s. BDs could also form in protoplanetary disks of stars \citep[e.g.][]{stamatellos09,kratter10,vorobyov13}. \citet{basu12} showed that clumps in protoplanetary disks could potentially be ejected by internal dynamical interactions between the clumps, and they would be ejected with lower velocities ($\sim$0.8 km/s), but these velocities would still kick these BD clumps far from their original positions by the age of NGC 2244. Ejection velocities below 0.3 km/s are needed in order to keep BDs formed by any of these mechanisms close to the their birth position. Overall, these formation scenarios of BDs do not provide a successful explanation for the observed spatial distribution of BDs either.

In order to further study the nature of this finding we need to study a larger region, and confirm the membership of a larger sample of BDs. If this behavior extends to other OB stars of the cluster we would be able to confirm whether we are indeed seeing the effect of OB stars on the formation of BDs.

We found the disk fraction of NGC 2244 to be 39$\pm$9\%. This value is lower than the disk fraction found in nearby SFRs of similar ages, although it is still within errors with the disk fraction of Taurus or IC 348. The low prevalence of disks in NGC 2244 was already reported in \citet{balog07}, where the disk fraction was also found to decrease when considering only the members closest to the O-type stars. The disk depletion timescale sets the time that is available for the formation of planets, and it is therefore a fundamental parameter in order to understand the formation of planets. The low fraction of NGC 2244 members that still have protoplanetary disk can be a consequence of faster disk dispersal due to the presence of OB stars. 

Overall, we found that the OB stars in NGC 2244 may have had an important role in the early evolution of the cluster, by potentially increasing the efficiency of BD formation and decreasing the lifetime of protoplanetary disks around both stars and BDs members of the cluster.

\begin{acknowledgements}
   V.A-A., K.M. and K.K. acknowledge funding by the Science and Technology Foundation of Portugal (FCT), grants No. IF/00194/2015, PTDC/FIS-AST/28731/2017, UIDB/00099/2020, SFRH/BD/143433/2019 and PTDC/FIS-AST/7002/2020. R.S. acknowledges financial support from the State Agency for Research of the Spanish MCIU through the "Center of Excellence Severo Ochoa" award for the Instituto de Astrof\'isica de Andaluc\'ia (SEV-2017-0709) and financial support from national project PGC2018-095049-B-C21 (MCIU/AEI/FEDER, UE).
\end{acknowledgements}

\bibliographystyle{aa} 
\bibliography{kmos}

\begin{thebibliography}{107}
\expandafter\ifx\csname natexlab\endcsname\relax\def\natexlab#1{#1}\fi

\bibitem[{{Allard} {et~al.}(2012){Allard}, {Homeier}, \& {Freytag}}]{allard12}
{Allard}, F., {Homeier}, D., \& {Freytag}, B. 2012, Philosophical Transactions
  of the Royal Society of London Series A, 370, 2765

\bibitem[{{Allers} \& {Liu}(2013)}]{allers13}
{Allers}, K.~N. \& {Liu}, M.~C. 2013, \apj, 772, 79

\bibitem[{{Almendros-Abad} {et~al.}(2022){Almendros-Abad}, {Mu{\v{z}}i{\'c}},
  {Moitinho}, {Krone-Martins}, \& {Kubiak}}]{almendros22}
{Almendros-Abad}, V., {Mu{\v{z}}i{\'c}}, K., {Moitinho}, A., {Krone-Martins},
  A., \& {Kubiak}, K. 2022, \aap, 657, A129

\bibitem[{{Alves de Oliveira} {et~al.}(2012){Alves de Oliveira}, {Moraux},
  {Bouvier}, \& {Bouy}}]{alvesdeoliveira12}
{Alves de Oliveira}, C., {Moraux}, E., {Bouvier}, J., \& {Bouy}, H. 2012, \aap,
  539, A151

\bibitem[{{Andersen} {et~al.}(2008){Andersen}, {Meyer}, {Greissl}, \&
  {Aversa}}]{andersen08}
{Andersen}, M., {Meyer}, M.~R., {Greissl}, J., \& {Aversa}, A. 2008, \apjl,
  683, L183

\bibitem[{{Balog} {et~al.}(2007){Balog}, {Muzerolle}, {Rieke}, {Su}, {Young},
  \& {Megeath}}]{balog07}
{Balog}, Z., {Muzerolle}, J., {Rieke}, G.~H., {et~al.} 2007, \apj, 660, 1532

\bibitem[{{Baraffe} {et~al.}(2015){Baraffe}, {Homeier}, {Allard}, \&
  {Chabrier}}]{baraffe15}
{Baraffe}, I., {Homeier}, D., {Allard}, F., \& {Chabrier}, G. 2015, \aap, 577,
  A42

\bibitem[{{Barrado y Navascu{\'e}s} \& {Mart{\'\i}n}(2003)}]{barrado03}
{Barrado y Navascu{\'e}s}, D. \& {Mart{\'\i}n}, E.~L. 2003, \aj, 126, 2997

\bibitem[{{Basu} \& {Vorobyov}(2012)}]{basu12}
{Basu}, S. \& {Vorobyov}, E.~I. 2012, \apj, 750, 30

\bibitem[{{Bate}(2012)}]{bate12}
{Bate}, M.~R. 2012, \mnras, 419, 3115

\bibitem[{{Bayo} {et~al.}(2011){Bayo}, {Barrado}, {Stauffer},
  {Morales-Calder{\'o}n}, {Melo}, {Hu{\'e}lamo}, {Bouy}, {Stelzer}, {Tamura},
  \& {Jayawardhana}}]{bayo11}
{Bayo}, A., {Barrado}, D., {Stauffer}, J., {et~al.} 2011, \aap, 536, A63

\bibitem[{{Bayo} {et~al.}(2008){Bayo}, {Rodrigo}, {Barrado Y Navascu{\'e}s},
  {Solano}, {Guti{\'e}rrez}, {Morales-Calder{\'o}n}, \& {Allard}}]{bayo08}
{Bayo}, A., {Rodrigo}, C., {Barrado Y Navascu{\'e}s}, D., {et~al.} 2008, \aap,
  492, 277

\bibitem[{{Bell} {et~al.}(2013){Bell}, {Naylor}, {Mayne}, {Jeffries}, \&
  {Littlefair}}]{bell13}
{Bell}, C. P.~M., {Naylor}, T., {Mayne}, N.~J., {Jeffries}, R.~D., \&
  {Littlefair}, S.~P. 2013, \mnras, 434, 806

\bibitem[{{Bertin} \& {Arnouts}(1996)}]{bertin96}
{Bertin}, E. \& {Arnouts}, S. 1996, \aaps, 117, 393

\bibitem[{{Bonnell} {et~al.}(2008){Bonnell}, {Clark}, \& {Bate}}]{bonnell08}
{Bonnell}, I.~A., {Clark}, P., \& {Bate}, M.~R. 2008, \mnras, 389, 1556

\bibitem[{{Bressan} {et~al.}(2012){Bressan}, {Marigo}, {Girardi}, {Salasnich},
  {Dal Cero}, {Rubele}, \& {Nanni}}]{bressan12}
{Bressan}, A., {Marigo}, P., {Girardi}, L., {et~al.} 2012, \mnras, 427, 127

\bibitem[{{Cardelli} {et~al.}(1989){Cardelli}, {Clayton}, \&
  {Mathis}}]{cardelli89}
{Cardelli}, J.~A., {Clayton}, G.~C., \& {Mathis}, J.~S. 1989, \apj, 345, 245

\bibitem[{{Chabrier}(2003)}]{chabrier03}
{Chabrier}, G. 2003, \pasp, 115, 763

\bibitem[{{Chabrier}(2005)}]{chabrier05}
{Chabrier}, G. 2005, in Astrophysics and Space Science Library, Vol. 327, The
  Initial Mass Function 50 Years Later, ed. E.~{Corbelli}, F.~{Palla}, \&
  H.~{Zinnecker}, 41

\bibitem[{{Chabrier} {et~al.}(2022){Chabrier}, {Baraffe}, {Phillips}, \&
  {Debras}}]{chabrier22}
{Chabrier}, G., {Baraffe}, I., {Phillips}, M., \& {Debras}, F. 2022, arXiv
  e-prints, arXiv:2212.07153

\bibitem[{{Chen} {et~al.}(2007){Chen}, {de Grijs}, \& {Zhao}}]{chen07}
{Chen}, L., {de Grijs}, R., \& {Zhao}, J.~L. 2007, \aj, 134, 1368

\bibitem[{{Cieza} {et~al.}(2007){Cieza}, {Padgett}, {Stapelfeldt}, {Augereau},
  {Harvey}, {Evans}, {Mer{\'\i}n}, {Koerner}, {Sargent}, {van Dishoeck},
  {Allen}, {Blake}, {Brooke}, {Chapman}, {Huard}, {Lai}, {Mundy}, {Myers},
  {Spiesman}, \& {Wahhaj}}]{cieza07}
{Cieza}, L., {Padgett}, D.~L., {Stapelfeldt}, K.~R., {et~al.} 2007, \apj, 667,
  308

\bibitem[{{Crapsi} {et~al.}(2008){Crapsi}, {van Dishoeck}, {Hogerheijde},
  {Pontoppidan}, \& {Dullemond}}]{crapsi08}
{Crapsi}, A., {van Dishoeck}, E.~F., {Hogerheijde}, M.~R., {Pontoppidan},
  K.~M., \& {Dullemond}, C.~P. 2008, \aap, 486, 245

\bibitem[{{Cushing} {et~al.}(2005){Cushing}, {Rayner}, \& {Vacca}}]{cushing05}
{Cushing}, M.~C., {Rayner}, J.~T., \& {Vacca}, W.~D. 2005, \apj, 623, 1115

\bibitem[{{Damian} {et~al.}(2021){Damian}, {Jose}, {Samal}, {Moraux}, {Das}, \&
  {Patra}}]{damian21}
{Damian}, B., {Jose}, J., {Samal}, M.~R., {et~al.} 2021, \mnras, 504, 2557

\bibitem[{{Davies} {et~al.}(2013){Davies}, {Agudo Berbel}, {Wiezorrek},
  {Cirasuolo}, {F{\"o}rster Schreiber}, {Jung}, {Muschielok}, {Ott}, {Ramsay},
  {Schlichter}, {Sharples}, \& {Wegner}}]{davies13}
{Davies}, R.~I., {Agudo Berbel}, A., {Wiezorrek}, E., {et~al.} 2013, \aap, 558,
  A56

\bibitem[{{Dobbie} {et~al.}(2002){Dobbie}, {Pinfield}, {Jameson}, \&
  {Hodgkin}}]{dobbie02}
{Dobbie}, P.~D., {Pinfield}, D.~J., {Jameson}, R.~F., \& {Hodgkin}, S.~T. 2002,
  \mnras, 335, L79

\bibitem[{{Esplin} \& {Luhman}(2019)}]{esplin19}
{Esplin}, T.~L. \& {Luhman}, K.~L. 2019, \aj, 158, 54

\bibitem[{{Esplin} \& {Luhman}(2020)}]{esplin20}
{Esplin}, T.~L. \& {Luhman}, K.~L. 2020, \aj, 159, 282

\bibitem[{{Esplin} \& {Luhman}(2022)}]{esplin22}
{Esplin}, T.~L. \& {Luhman}, K.~L. 2022, \aj, 163, 64

\bibitem[{{Esplin} {et~al.}(2017){Esplin}, {Luhman}, {Faherty}, {Mamajek}, \&
  {Bochanski}}]{esplin17}
{Esplin}, T.~L., {Luhman}, K.~L., {Faherty}, J.~K., {Mamajek}, E.~E., \&
  {Bochanski}, J.~J. 2017, \aj, 154, 46

\bibitem[{{Esplin} {et~al.}(2018){Esplin}, {Luhman}, {Miller}, \&
  {Mamajek}}]{esplin18}
{Esplin}, T.~L., {Luhman}, K.~L., {Miller}, E.~B., \& {Mamajek}, E.~E. 2018,
  \aj, 156, 75

\bibitem[{{Fang} {et~al.}(2009){Fang}, {van Boekel}, {Wang}, {Carmona},
  {Sicilia-Aguilar}, \& {Henning}}]{fang09}
{Fang}, M., {van Boekel}, R., {Wang}, W., {et~al.} 2009, \aap, 504, 461

\bibitem[{{Fazio} {et~al.}(2004){Fazio}, {Hora}, {Allen}, {Ashby}, {Barmby},
  {Deutsch}, {Huang}, {Kleiner}, {Marengo}, {Megeath}, {Melnick}, {Pahre},
  {Patten}, {Polizotti}, {Smith}, {Taylor}, {Wang}, {Willner}, {Hoffmann},
  {Pipher}, {Forrest}, {McMurty}, {McCreight}, {McKelvey}, {McMurray}, {Koch},
  {Moseley}, {Arendt}, {Mentzell}, {Marx}, {Losch}, {Mayman}, {Eichhorn},
  {Krebs}, {Jhabvala}, {Gezari}, {Fixsen}, {Flores}, {Shakoorzadeh}, {Jungo},
  {Hakun}, {Workman}, {Karpati}, {Kichak}, {Whitley}, {Mann}, {Tollestrup},
  {Eisenhardt}, {Stern}, {Gorjian}, {Bhattacharya}, {Carey}, {Nelson},
  {Glaccum}, {Lacy}, {Lowrance}, {Laine}, {Reach}, {Stauffer}, {Surace},
  {Wilson}, {Wright}, {Hoffman}, {Domingo}, \& {Cohen}}]{fazio04}
{Fazio}, G.~G., {Hora}, J.~L., {Allen}, L.~E., {et~al.} 2004, \apjs, 154, 10

\bibitem[{{Fedele} {et~al.}(2010){Fedele}, {van den Ancker}, {Henning},
  {Jayawardhana}, \& {Oliveira}}]{fedele10}
{Fedele}, D., {van den Ancker}, M.~E., {Henning}, T., {Jayawardhana}, R., \&
  {Oliveira}, J.~M. 2010, \aap, 510, A72

\bibitem[{{Fernandes} {et~al.}(2012){Fernandes}, {Gregorio-Hetem}, \&
  {Hetem}}]{fernandes12}
{Fernandes}, B., {Gregorio-Hetem}, J., \& {Hetem}, A. 2012, \aap, 541, A95

\bibitem[{{Freudling} {et~al.}(2013){Freudling}, {Romaniello}, {Bramich},
  {Ballester}, {Forchi}, {Garc{\'{\i}}a-Dabl{\'o}}, {Moehler}, \&
  {Neeser}}]{freudling13}
{Freudling}, W., {Romaniello}, M., {Bramich}, D.~M., {et~al.} 2013, \aap, 559,
  A96

\bibitem[{{Gahm} {et~al.}(2007){Gahm}, {Grenman}, {Fredriksson}, \&
  {Kristen}}]{gahm07}
{Gahm}, G.~F., {Grenman}, T., {Fredriksson}, S., \& {Kristen}, H. 2007, \aj,
  133, 1795

\bibitem[{{Gahm} {et~al.}(2013){Gahm}, {Persson}, {M{\"a}kel{\"a}}, \&
  {Haikala}}]{gahm13}
{Gahm}, G.~F., {Persson}, C.~M., {M{\"a}kel{\"a}}, M.~M., \& {Haikala}, L.~K.
  2013, \aap, 555, A57

\bibitem[{{Geha} {et~al.}(2013){Geha}, {Brown}, {Tumlinson}, {Kalirai},
  {Simon}, {Kirby}, {VandenBerg}, {Mu{\~n}oz}, {Avila}, {Guhathakurta}, \&
  {Ferguson}}]{geha13}
{Geha}, M., {Brown}, T.~M., {Tumlinson}, J., {et~al.} 2013, \apj, 771, 29

\bibitem[{{Gennaro} \& {Robberto}(2020)}]{gennaro20}
{Gennaro}, M. \& {Robberto}, M. 2020, \apj, 896, 80

\bibitem[{{Gorlova} {et~al.}(2003){Gorlova}, {Meyer}, {Rieke}, \&
  {Liebert}}]{gorlova03}
{Gorlova}, N.~I., {Meyer}, M.~R., {Rieke}, G.~H., \& {Liebert}, J. 2003, \apj,
  593, 1074

\bibitem[{{Grenman} \& {Gahm}(2014)}]{grenman14}
{Grenman}, T. \& {Gahm}, G.~F. 2014, \aap, 565, A107

\bibitem[{{Gutermuth} {et~al.}(2009){Gutermuth}, {Megeath}, {Myers}, {Allen},
  {Pipher}, \& {Fazio}}]{gutermuth09}
{Gutermuth}, R.~A., {Megeath}, S.~T., {Myers}, P.~C., {et~al.} 2009, \apjs,
  184, 18

\bibitem[{{Haisch} {et~al.}(2001){Haisch}, {Lada}, \& {Lada}}]{haisch01}
{Haisch}, Karl~E., J., {Lada}, E.~A., \& {Lada}, C.~J. 2001, \apjl, 553, L153

\bibitem[{{Haworth} {et~al.}(2015){Haworth}, {Facchini}, \&
  {Clarke}}]{haworth15}
{Haworth}, T.~J., {Facchini}, S., \& {Clarke}, C.~J. 2015, \mnras, 446, 1098

\bibitem[{{Hensberge} {et~al.}(2000){Hensberge}, {Pavlovski}, \&
  {Verschueren}}]{hensberge00}
{Hensberge}, H., {Pavlovski}, K., \& {Verschueren}, W. 2000, \aap, 358, 553

\bibitem[{{Herczeg} \& {Hillenbrand}(2014)}]{herczeg14}
{Herczeg}, G.~J. \& {Hillenbrand}, L.~A. 2014, \apj, 786, 97

\bibitem[{{Hosek} {et~al.}(2019){Hosek}, {Lu}, {Anderson}, {Najarro}, {Ghez},
  {Morris}, {Clarkson}, \& {Albers}}]{hosek19}
{Hosek}, Matthew~W., J., {Lu}, J.~R., {Anderson}, J., {et~al.} 2019, \apj, 870,
  44

\bibitem[{{Johnstone} {et~al.}(1998){Johnstone}, {Hollenbach}, \&
  {Bally}}]{johnstone98}
{Johnstone}, D., {Hollenbach}, D., \& {Bally}, J. 1998, \apj, 499, 758

\bibitem[{{Jones} \& {Bate}(2018)}]{jones18}
{Jones}, M.~O. \& {Bate}, M.~R. 2018, \mnras, 478, 2650

\bibitem[{{Kirkpatrick} {et~al.}(2006){Kirkpatrick}, {Barman}, {Burgasser},
  {McGovern}, {McLean}, {Tinney}, \& {Lowrance}}]{kirkpatrick06}
{Kirkpatrick}, J.~D., {Barman}, T.~S., {Burgasser}, A.~J., {et~al.} 2006, \apj,
  639, 1120

\bibitem[{{Kratter} {et~al.}(2010){Kratter}, {Matzner}, {Krumholz}, \&
  {Klein}}]{kratter10}
{Kratter}, K.~M., {Matzner}, C.~D., {Krumholz}, M.~R., \& {Klein}, R.~I. 2010,
  \apj, 708, 1585

\bibitem[{{Kroupa}(2001)}]{kroupa01}
{Kroupa}, P. 2001, \mnras, 322, 231

\bibitem[{{Kroupa}(2002)}]{kroupa02}
{Kroupa}, P. 2002, Science, 295, 82

\bibitem[{{Li} {et~al.}(2023){Li}, {Liu}, {Zhang}, {Tian}, {Fu}, {Li}, \&
  {Yan}}]{jiadong23}
{Li}, J., {Liu}, C., {Zhang}, Z.-Y., {et~al.} 2023, \nat, 613, 460

\bibitem[{{Lodieu}(2013)}]{lodieu13}
{Lodieu}, N. 2013, \mnras, 431, 3222

\bibitem[{{Lucas} {et~al.}(2001){Lucas}, {Roche}, {Allard}, \&
  {Hauschildt}}]{lucas01}
{Lucas}, P.~W., {Roche}, P.~F., {Allard}, F., \& {Hauschildt}, P.~H. 2001,
  \mnras, 326, 695

\bibitem[{{Luhman}(2007)}]{luhman07}
{Luhman}, K.~L. 2007, \apjs, 173, 104

\bibitem[{{Luhman}(2012)}]{luhman12}
{Luhman}, K.~L. 2012, \araa, 50, 65

\bibitem[{{Luhman} {et~al.}(2008){Luhman}, {Allen}, {Allen}, {Gutermuth},
  {Hartmann}, {Mamajek}, {Megeath}, {Myers}, \& {Fazio}}]{luhman08}
{Luhman}, K.~L., {Allen}, L.~E., {Allen}, P.~R., {et~al.} 2008, \apj, 675, 1375

\bibitem[{{Luhman} {et~al.}(2003){Luhman}, {Brice{\~n}o}, {Stauffer},
  {Hartmann}, {Barrado y Navascu{\'e}s}, \& {Caldwell}}]{luhman03a}
{Luhman}, K.~L., {Brice{\~n}o}, C., {Stauffer}, J.~R., {et~al.} 2003, \apj,
  590, 348

\bibitem[{{Luhman} {et~al.}(2017){Luhman}, {Mamajek}, {Shukla}, \&
  {Loutrel}}]{luhman17}
{Luhman}, K.~L., {Mamajek}, E.~E., {Shukla}, S.~J., \& {Loutrel}, N.~P. 2017,
  \aj, 153, 46

\bibitem[{{Luo} {et~al.}(2015){Luo}, {Zhao}, {Zhao}, {Deng}, {Liu}, {Jing},
  {Wang}, {Zhang}, {Shi}, {Cui}, {Chu}, {Li}, {Bai}, {Wu}, {Cai}, {Cao}, {Cao},
  {Carlin}, {Chen}, {Chen}, {Chen}, {Chen}, {Chen}, {Chen}, {Chen},
  {Christlieb}, {Chu}, {Cui}, {Dong}, {Du}, {Fan}, {Feng}, {Fu}, {Gao}, {Gong},
  {Gu}, {Guo}, {Han}, {He}, {Hou}, {Hou}, {Hou}, {Hu}, {Hu}, {Hu}, {Huo},
  {Jia}, {Jiang}, {Jiang}, {Jiang}, {Jin}, {Kong}, {Kong}, {Lei}, {Li}, {Li},
  {Li}, {Li}, {Li}, {Li}, {Li}, {Li}, {Li}, {Li}, {Li}, {Li}, {Liang}, {Lin},
  {Liu}, {Liu}, {Liu}, {Liu}, {Lu}, {Luo}, {Mao}, {Newberg}, {Ni}, {Qi}, {Qi},
  {Shen}, {Shi}, {Song}, {Song}, {Su}, {Su}, {Tang}, {Tao}, {Tian}, {Wang},
  {Wang}, {Wang}, {Wang}, {Wang}, {Wang}, {Wang}, {Wang}, {Wang}, {Wang},
  {Wang}, {Wang}, {Wang}, {Wang}, {Wang}, {Wang}, {Wang}, {Wang}, {Wang},
  {Wang}, {Wei}, {Wei}, {Wu}, {Wu}, {Wu}, {Wu}, {Xing}, {Xu}, {Xu}, {Xu},
  {Yan}, {Yang}, {Yang}, {Yang}, {Yang}, {Yao}, {Yu}, {Yuan}, {Yuan}, {Yuan},
  {Yuan}, {Zhai}, {Zhang}, {Zhang}, {Zhang}, {Zhang}, {Zhang}, {Zhang},
  {Zhang}, {Zhang}, {Zhao}, {Zhou}, {Zhou}, {Zhu}, {Zhu}, {Zou}, \&
  {Zuo}}]{luo15}
{Luo}, A.~L., {Zhao}, Y.-H., {Zhao}, G., {et~al.} 2015, Research in Astronomy
  and Astrophysics, 15, 1095

\bibitem[{{M{\"a}kel{\"a}} {et~al.}(2014){M{\"a}kel{\"a}}, {Haikala}, \&
  {Gahm}}]{makela14}
{M{\"a}kel{\"a}}, M.~M., {Haikala}, L.~K., \& {Gahm}, G.~F. 2014, \aap, 567,
  A108

\bibitem[{{Meng} {et~al.}(2017){Meng}, {Rieke}, {Su}, \&
  {G{\'a}sp{\'a}r}}]{meng17}
{Meng}, H. Y.~A., {Rieke}, G.~H., {Su}, K. Y.~L., \& {G{\'a}sp{\'a}r}, A. 2017,
  \apj, 836, 34

\bibitem[{{Michel} {et~al.}(2021){Michel}, {van der Marel}, \&
  {Matthews}}]{michel21}
{Michel}, A., {van der Marel}, N., \& {Matthews}, B.~C. 2021, \apj, 921, 72

\bibitem[{{Mu{\v{z}}i{\'c}} {et~al.}(2022){Mu{\v{z}}i{\'c}}, {Almendros-Abad},
  {Bouy}, {Kubiak}, {Pe{\~n}a Ram{\'\i}rez}, {Krone-Martins}, {Moitinho}, \&
  {Concei{\c{c}}{\~a}o}}]{muzic22}
{Mu{\v{z}}i{\'c}}, K., {Almendros-Abad}, V., {Bouy}, H., {et~al.} 2022, \aap,
  668, A19

\bibitem[{{Mu{\v{z}}i{\'c}} {et~al.}(2017){Mu{\v{z}}i{\'c}}, {Sch{\"o}del},
  {Scholz}, {Geers}, {Jayawardhana}, {Ascenso}, \& {Cieza}}]{muzic17}
{Mu{\v{z}}i{\'c}}, K., {Sch{\"o}del}, R., {Scholz}, A., {et~al.} 2017, \mnras,
  471, 3699

\bibitem[{{Mu{\v{z}}i{\'c}} {et~al.}(2015){Mu{\v{z}}i{\'c}}, {Scholz}, {Geers},
  \& {Jayawardhana}}]{muzic15}
{Mu{\v{z}}i{\'c}}, K., {Scholz}, A., {Geers}, V.~C., \& {Jayawardhana}, R.
  2015, \apj, 810, 159

\bibitem[{{Mu{\v{z}}i{\'c}} {et~al.}(2019){Mu{\v{z}}i{\'c}}, {Scholz},
  {Pe{\~n}a Ram{\'\i}rez}, {Jayawardhana}, {Sch{\"o}del}, {Geers}, {Cieza}, \&
  {Bayo}}]{muzic19}
{Mu{\v{z}}i{\'c}}, K., {Scholz}, A., {Pe{\~n}a Ram{\'\i}rez}, K., {et~al.}
  2019, \apj, 881, 79

\bibitem[{{Noll} {et~al.}(2014){Noll}, {Kausch}, {Kimeswenger}, {Barden},
  {Jones}, {Modigliani}, {Szyszka}, \& {Taylor}}]{noll14}
{Noll}, S., {Kausch}, W., {Kimeswenger}, S., {et~al.} 2014, \aap, 567, A25

\bibitem[{{O'dell} {et~al.}(1993){O'dell}, {Wen}, \& {Hu}}]{odell93}
{O'dell}, C.~R., {Wen}, Z., \& {Hu}, X. 1993, \apj, 410, 696

\bibitem[{{Padoan} \& {Nordlund}(2004)}]{padoan04}
{Padoan}, P. \& {Nordlund}, {\r{A}}. 2004, \apj, 617, 559

\bibitem[{{Pe{\~n}a Ram{\'\i}rez} {et~al.}(2012){Pe{\~n}a Ram{\'\i}rez},
  {B{\'e}jar}, {Zapatero Osorio}, {Petr-Gotzens}, \& {Mart{\'\i}n}}]{pena12}
{Pe{\~n}a Ram{\'\i}rez}, K., {B{\'e}jar}, V.~J.~S., {Zapatero Osorio}, M.~R.,
  {Petr-Gotzens}, M.~G., \& {Mart{\'\i}n}, E.~L. 2012, \apj, 754, 30

\bibitem[{{Pecaut} \& {Mamajek}(2013)}]{pecaut13}
{Pecaut}, M.~J. \& {Mamajek}, E.~E. 2013, \apjs, 208, 9

\bibitem[{{Pfalzner} {et~al.}(2022){Pfalzner}, {Dehghani}, \&
  {Michel}}]{pfalzner22}
{Pfalzner}, S., {Dehghani}, S., \& {Michel}, A. 2022, \apjl, 939, L10

\bibitem[{{Poulton} {et~al.}(2008){Poulton}, {Robitaille}, {Greaves},
  {Bonnell}, {Williams}, \& {Heyer}}]{poulton08}
{Poulton}, C.~J., {Robitaille}, T.~P., {Greaves}, J.~S., {et~al.} 2008, \mnras,
  384, 1249

\bibitem[{{Preibisch} \& {Mamajek}(2008)}]{preibisch08}
{Preibisch}, T. \& {Mamajek}, E. 2008, {The Nearest OB Association:
  Scorpius-Centaurus (Sco OB2)}, ed. B.~{Reipurth}, Vol.~5, 235

\bibitem[{{Randich} {et~al.}(2022){Randich}, {Gilmore}, {Magrini}, {Sacco},
  {Jackson}, {Jeffries}, {Worley}, {Hourihane}, {Gonneau}, {Viscasillas
  Vazquez}, {Franciosini}, {Lewis}, {Alfaro}, {Allende Prieto}, {Bensby},
  {Blomme}, {Bragaglia}, {Flaccomio}, {Fran{\c{c}}ois}, {Irwin}, {Koposov},
  {Korn}, {Lanzafame}, {Pancino}, {Recio-Blanco}, {Smiljanic}, {Van Eck},
  {Zwitter}, {Asplund}, {Bonifacio}, {Feltzing}, {Binney}, {Drew}, {Ferguson},
  {Micela}, {Negueruela}, {Prusti}, {Rix}, {Vallenari}, {Bayo}, {Bergemann},
  {Biazzo}, {Carraro}, {Casey}, {Damiani}, {Frasca}, {Heiter}, {Hill},
  {Jofr{\'e}}, {de Laverny}, {Lind}, {Marconi}, {Martayan}, {Masseron},
  {Monaco}, {Morbidelli}, {Prisinzano}, {Sbordone}, {Sousa}, {Zaggia},
  {Adibekyan}, {Bonito}, {Caffau}, {Daflon}, {Feuillet}, {Gebran}, {Gonzalez
  Hernandez}, {Guiglion}, {Herrero}, {Lobel}, {Maiz Apellaniz}, {Merle},
  {Mikolaitis}, {Montes}, {Morel}, {Soubiran}, {Spina}, {Tabernero},
  {Tautvai{\v{s}}iene}, {Traven}, {Valentini}, {Van der Swaelmen}, {Villanova},
  {Wright}, {Abbas}, {Aguirre B{\o}rsen-Koch}, {Alves}, {Balaguer-Nunez},
  {Barklem}, {Barrado}, {Berlanas}, {Binks}, {Bressan}, {Capuzzo-Dolcetta},
  {Casagrande}, {Casamiquela}, {Collins}, {D'Orazi}, {Dantas}, {Debattista},
  {Delgado-Mena}, {Di Marcantonio}, {Drazdauskas}, {Evans}, {Famaey},
  {Franchini}, {Fr{\'e}mat}, {Friel}, {Fu}, {Geisler}, {Gerhard}, {Gonzalez
  Solares}, {Grebel}, {Gutierrez Albarran}, {Hatzidimitriou}, {Held},
  {Jim{\'e}nez-Esteban}, {J{\"o}nsson}, {Jordi}, {Khachaturyants},
  {Kordopatis}, {Kos}, {Lagarde}, {Mahy}, {Mapelli}, {Marfil}, {Martell},
  {Messina}, {Miglio}, {Minchev}, {Moitinho}, {Montalban}, {Monteiro},
  {Morossi}, {Mowlavi}, {Mucciarelli}, {Murphy}, {Nardetto}, {Ortolani},
  {Paletou}, {Palou{\v{s}}}, {Paunzen}, {Pickering}, {Quirrenbach}, {Re
  Fiorentin}, {Read}, {Romano}, {Ryde}, {Sanna}, {Santos}, {Seabroke},
  {Spagna}, {Steinmetz}, {Stonkut{\'e}}, {Sutorius}, {Th{\'e}venin}, {Tosi},
  {Tsantaki}, {Vink}, {Wright}, {Wyse}, {Zoccali}, {Zorec}, {Zucker}, \&
  {Walton}}]{randich22}
{Randich}, S., {Gilmore}, G., {Magrini}, L., {et~al.} 2022, \aap, 666, A121

\bibitem[{{Reipurth} \& {Clarke}(2001)}]{reipurth01}
{Reipurth}, B. \& {Clarke}, C. 2001, \aj, 122, 432

\bibitem[{{Reyl{\'e}} \& {Robin}(2001)}]{reyle01}
{Reyl{\'e}}, C. \& {Robin}, A.~C. 2001, \aap, 373, 886

\bibitem[{{Ribas} {et~al.}(2015){Ribas}, {Bouy}, \& {Mer{\'\i}n}}]{ribas15}
{Ribas}, {\'A}., {Bouy}, H., \& {Mer{\'\i}n}, B. 2015, \aap, 576, A52

\bibitem[{{Rom{\'a}n-Z{\'u}{\~n}iga} {et~al.}(2023){Rom{\'a}n-Z{\'u}{\~n}iga},
  {Kounkel}, {Hern{\'a}ndez}, {Pe{\~n}a Ram{\'\i}rez}, {L{\'o}pez-Valdivia},
  {Covey}, {Stutz}, {Roman-Lopes}, {Campbell}, {Khilfeh}, {Tapia},
  {Stringfellow}, {Downes}, {Stassun}, {Minniti}, {Bayo}, {Kim}, {Su{\'a}rez},
  {Ybarra}, {Fern{\'a}ndez-Trincado}, {Longa-Pe{\~n}a},
  {Ram{\'\i}rez-Preciado}, {Serna}, {Lane}, {Garc{\'\i}a-Hern{\'a}ndez},
  {Beaton}, {Bizyaev}, \& {Pan}}]{roman23}
{Rom{\'a}n-Z{\'u}{\~n}iga}, C.~G., {Kounkel}, M., {Hern{\'a}ndez}, J., {et~al.}
  2023, \aj, 165, 51

\bibitem[{{Rom{\'a}n-Z{\'u}{\~n}iga} \& {Lada}(2008)}]{sfh_rosette}
{Rom{\'a}n-Z{\'u}{\~n}iga}, C.~G. \& {Lada}, E.~A. 2008, in Handbook of Star
  Forming Regions, Volume I, ed. B.~{Reipurth}, Vol.~4, 928

\bibitem[{{Salpeter}(1955)}]{salpeter55}
{Salpeter}, E.~E. 1955, \apj, 121, 161

\bibitem[{{Scholz} {et~al.}(2013){Scholz}, {Geers}, {Clark}, {Jayawardhana}, \&
  {Muzic}}]{scholz13}
{Scholz}, A., {Geers}, V., {Clark}, P., {Jayawardhana}, R., \& {Muzic}, K.
  2013, \apj, 775, 138

\bibitem[{{Scholz} {et~al.}(2012){Scholz}, {Muzic}, {Geers}, {Bonavita},
  {Jayawardhana}, \& {Tamura}}]{scholz12a}
{Scholz}, A., {Muzic}, K., {Geers}, V., {et~al.} 2012, \apj, 744, 6

\bibitem[{{Sharples} {et~al.}(2013){Sharples}, {Bender}, {Agudo Berbel},
  {Bezawada}, {Castillo}, {Cirasuolo}, {Davidson}, {Davies}, {Dubbeldam},
  {Fairley}, {Finger}, {F{\"o}rster Schreiber}, {Gonte}, {Hess}, {Jung},
  {Lewis}, {Lizon}, {Muschielok}, {Pasquini}, {Pirard}, {Popovic}, {Ramsay},
  {Rees}, {Richter}, {Riquelme}, {Rodrigues}, {Saviane}, {Schlichter},
  {Schmidtobreick}, {Segovia}, {Smette}, {Szeifert}, {van Kesteren}, {Wegner},
  \& {Wiezorrek}}]{sharpies13}
{Sharples}, R., {Bender}, R., {Agudo Berbel}, A., {et~al.} 2013, The Messenger,
  151, 21

\bibitem[{{Slesnick} {et~al.}(2004){Slesnick}, {Hillenbrand}, \&
  {Carpenter}}]{slesnick04}
{Slesnick}, C.~L., {Hillenbrand}, L.~A., \& {Carpenter}, J.~M. 2004, \apj, 610,
  1045

\bibitem[{{Smette} {et~al.}(2015){Smette}, {Sana}, {Noll}, {Horst}, {Kausch},
  {Kimeswenger}, {Barden}, {Szyszka}, {Jones}, {Gallenne}, {Vinther},
  {Ballester}, \& {Taylor}}]{smette15}
{Smette}, A., {Sana}, H., {Noll}, S., {et~al.} 2015, \aap, 576, A77

\bibitem[{{Stamatellos} {et~al.}(2011){Stamatellos}, {Maury}, {Whitworth}, \&
  {Andr{\'e}}}]{stamatellos11}
{Stamatellos}, D., {Maury}, A., {Whitworth}, A., \& {Andr{\'e}}, P. 2011,
  \mnras, 413, 1787

\bibitem[{{Stamatellos} \& {Whitworth}(2009)}]{stamatellos09}
{Stamatellos}, D. \& {Whitworth}, A.~P. 2009, \mnras, 392, 413

\bibitem[{{Stolte} {et~al.}(2006){Stolte}, {Brandner}, {Brandl}, \&
  {Zinnecker}}]{stolte06}
{Stolte}, A., {Brandner}, W., {Brandl}, B., \& {Zinnecker}, H. 2006, \aj, 132,
  253

\bibitem[{{Su{\'a}rez} {et~al.}(2019){Su{\'a}rez}, {Downes},
  {Rom{\'a}n-Z{\'u}{\~n}iga}, {Cervi{\~n}o}, {Brice{\~n}o}, {Petr-Gotzens}, \&
  {Vivas}}]{suarez19}
{Su{\'a}rez}, G., {Downes}, J.~J., {Rom{\'a}n-Z{\'u}{\~n}iga}, C., {et~al.}
  2019, \mnras, 486, 1718

\bibitem[{{Testi} {et~al.}(2001){Testi}, {D'Antona}, {Ghinassi}, {Licandro},
  {Magazz{\`u}}, {Maiolino}, {Mannucci}, {Marconi}, {Nagar}, {Natta}, \&
  {Oliva}}]{testi01}
{Testi}, L., {D'Antona}, F., {Ghinassi}, F., {et~al.} 2001, \apjl, 552, L147

\bibitem[{{Vorobyov}(2013)}]{vorobyov13}
{Vorobyov}, E.~I. 2013, \aap, 552, A129

\bibitem[{{Wang} {et~al.}(2008){Wang}, {Townsley}, {Feigelson}, {Broos},
  {Getman}, {Rom{\'a}n-Z{\'u}{\~n}iga}, \& {Lada}}]{wang08}
{Wang}, J., {Townsley}, L.~K., {Feigelson}, E.~D., {et~al.} 2008, \apj, 675,
  464

\bibitem[{{Wareing} {et~al.}(2018){Wareing}, {Pittard}, {Wright}, \&
  {Falle}}]{wareing18}
{Wareing}, C.~J., {Pittard}, J.~M., {Wright}, N.~J., \& {Falle}, S.~A.~E.~G.
  2018, \mnras, 475, 3598

\bibitem[{{Weights} {et~al.}(2009){Weights}, {Lucas}, {Roche}, {Pinfield}, \&
  {Riddick}}]{weights09}
{Weights}, D.~J., {Lucas}, P.~W., {Roche}, P.~F., {Pinfield}, D.~J., \&
  {Riddick}, F. 2009, \mnras, 392, 817

\bibitem[{{Werner} {et~al.}(2004){Werner}, {Roellig}, {Low}, {Rieke}, {Rieke},
  {Hoffmann}, {Young}, {Houck}, {Brandl}, {Fazio}, {Hora}, {Gehrz}, {Helou},
  {Soifer}, {Stauffer}, {Keene}, {Eisenhardt}, {Gallagher}, {Gautier}, {Irace},
  {Lawrence}, {Simmons}, {Van Cleve}, {Jura}, {Wright}, \&
  {Cruikshank}}]{werner04}
{Werner}, M.~W., {Roellig}, T.~L., {Low}, F.~J., {et~al.} 2004, \apjs, 154, 1

\bibitem[{{Whitworth} \& {Zinnecker}(2004)}]{whitworth04}
{Whitworth}, A.~P. \& {Zinnecker}, H. 2004, \aap, 427, 299

\bibitem[{{Williams} \& {Cieza}(2011)}]{williams11}
{Williams}, J.~P. \& {Cieza}, L.~A. 2011, \araa, 49, 67

\bibitem[{{Winston} {et~al.}(2011){Winston}, {Wolk}, {Bourke}, {Megeath},
  {Gutermuth}, \& {Spitzbart}}]{winston11}
{Winston}, E., {Wolk}, S.~J., {Bourke}, T.~L., {et~al.} 2011, \apj, 743, 166

\bibitem[{{Winter} \& {Haworth}(2022)}]{winter22}
{Winter}, A.~J. \& {Haworth}, T.~J. 2022, European Physical Journal Plus, 137,
  1132

\bibitem[{{Wolk} {et~al.}(2008){Wolk}, {Bourke}, \& {Vigil}}]{wolk08}
{Wolk}, S.~J., {Bourke}, T.~L., \& {Vigil}, M. 2008, in Handbook of Star
  Forming Regions, Volume II, ed. B.~{Reipurth}, Vol.~5, 124

\bibitem[{{Yan} {et~al.}(2020){Yan}, {Jerabkova}, \& {Kroupa}}]{yan20}
{Yan}, Z., {Jerabkova}, T., \& {Kroupa}, P. 2020, \aap, 637, A68

\end{thebibliography}

\begin{appendix}

\section{Gravity-sensitive indices decision boundary}
\label{app_decision_boundary}

We perform a derivation of decision boundaries for the three gravity-sensitive indices used for the membership classification in Sect.~\ref{analysis_youth}. These boundaries give for each SpT bin, the spectral index value that separates young low-gravity members from non-members. We use the data set presented in \citet{almendros22}. We used the decision function given by the Support Vector Machine (SVM) method applied to the spectral index and SpT using the radial basis function kernel in order to define the decision boundary between the ``young'' and ``field'' classes of the \citet{almendros22} data set. SVM maximizes the distance between the different classes and a separating hyperplane defined on a feature space that is transformed using a kernel. We use the same methodology to optimize the hyper-parameters and to obtain the decision function as in \citet{almendros22}. The decision boundaries are given in Table~\ref{tab:tab_decision_boundary}.

\begin{table}
    \caption{Low-gravity decision boundary for the three inspected gravity-sensitive spectral indices: HPI \citep{scholz13}, TLI-g \citep{almendros22} and WH \citep{weights09}.}
    \begin{center}
        \begin{tabular}{l c c c}
            \hline\hline
            SpT & HPI & TLI-g & WH \\ 
            \hline
            M0 & 0.883 & 0.999 & 1.07 \\ 
            M0.5 & 0.892 & 1.003 & 1.067 \\ 
            M1 & 0.9 & 1.006 & 1.063 \\ 
            M1.5 & 0.909 & 1.009 & 1.059 \\ 
            M2 & 0.917 & 1.011 & 1.054 \\ 
            M2.5 & 0.924 & 1.012 & 1.048 \\ 
            M3 & 0.93 & 1.013 & 1.042 \\ 
            M3.5 & 0.936 & 1.013 & 1.035 \\ 
            M4 & 0.943 & 1.011 & 1.027 \\ 
            M4.5 & 0.952 & 1.01 & 1.018 \\ 
            M5 & 0.965 & 1.006 & 1.009 \\ 
            M5.5 & 0.985 & 1.002 & 0.999 \\ 
            M6 & 1.013 & 0.997 & 0.988 \\ 
            M6.5 & 1.049 & 0.991 & 0.977 \\ 
            M7 & 1.091 & 0.985 & 0.966 \\ 
            M7.5 & 1.134 & 0.978 & 0.954 \\ 
            M8 & 1.175 & 0.972 & 0.942 \\ 
            M8.5 & 1.211 & 0.966 & 0.93 \\ 
            M9 & 1.243 & 0.962 & 0.918 \\ 
            M9.5 & 1.274 & 0.958 & 0.906 \\ 
            L0 & 1.306 & 0.955 & 0.894 \\ 
            L0.5 & 1.341 & 0.953 & 0.884 \\ 
            L1 & 1.383 & 0.95 & 0.873 \\ 
            L1.5 & 1.439 & 0.948 & 0.864 \\ 
            L2 & 1.505 & 0.945 & 0.855 \\ 
            \hline 
        \end{tabular}
    \end{center}
    \label{tab:tab_decision_boundary}
\end{table}

\section{Photospheric infrared colors}
\label{app_excess}

\begin{figure*}[hbt!]
    \centering
    \includegraphics[width=\textwidth]{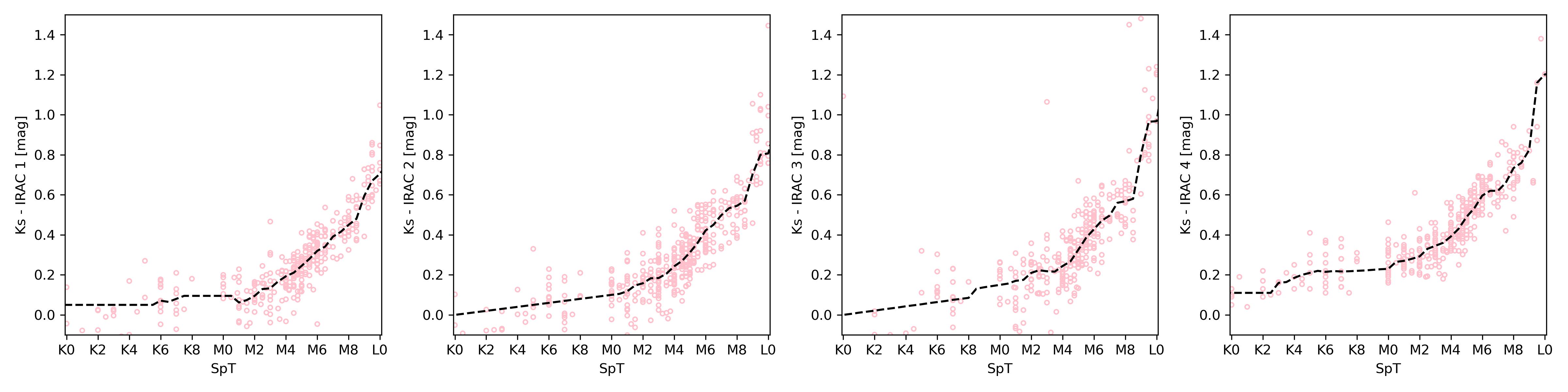}
    \caption{$K_{\mathrm{S}}$ - IRAC color as a function of SpT for a compilation of diskless young sources from the literature (see text) for the four IRAC channels. The black dashed line represents the mean photospheric colors for each SpT bin we derived.}
    \label{fig:app_ir_colors}
\end{figure*}

We derive the $K\mathrm{_S}$-IRAC photospheric colors for a compilation of known diskless members of the Cha-I \citep{luhman08,esplin17}, Upper Sco \citep{esplin18}, $\rho$-Ophiuchus \citep{esplin20}, Taurus \citep{esplin19} and Corona Australis \citep{esplin22} SFRs between K0 and L0. We restrict the analysis to low-extincted sources (Av$<$2 mag) and performed a running average with 1 SpT window for each 0.5 SpT (see Fig.~\ref{fig:app_ir_colors}). We provide in Table~\ref{tab:tab_app_irac_colors} the derived colors. These are the colors used in Sect.~\ref{analysis_excess} to obtain the infrared excess in each IRAC band for the NGC 2244 candidate members.

\begin{figure*}[hbt!]
    \centering
    \includegraphics[width=\textwidth]{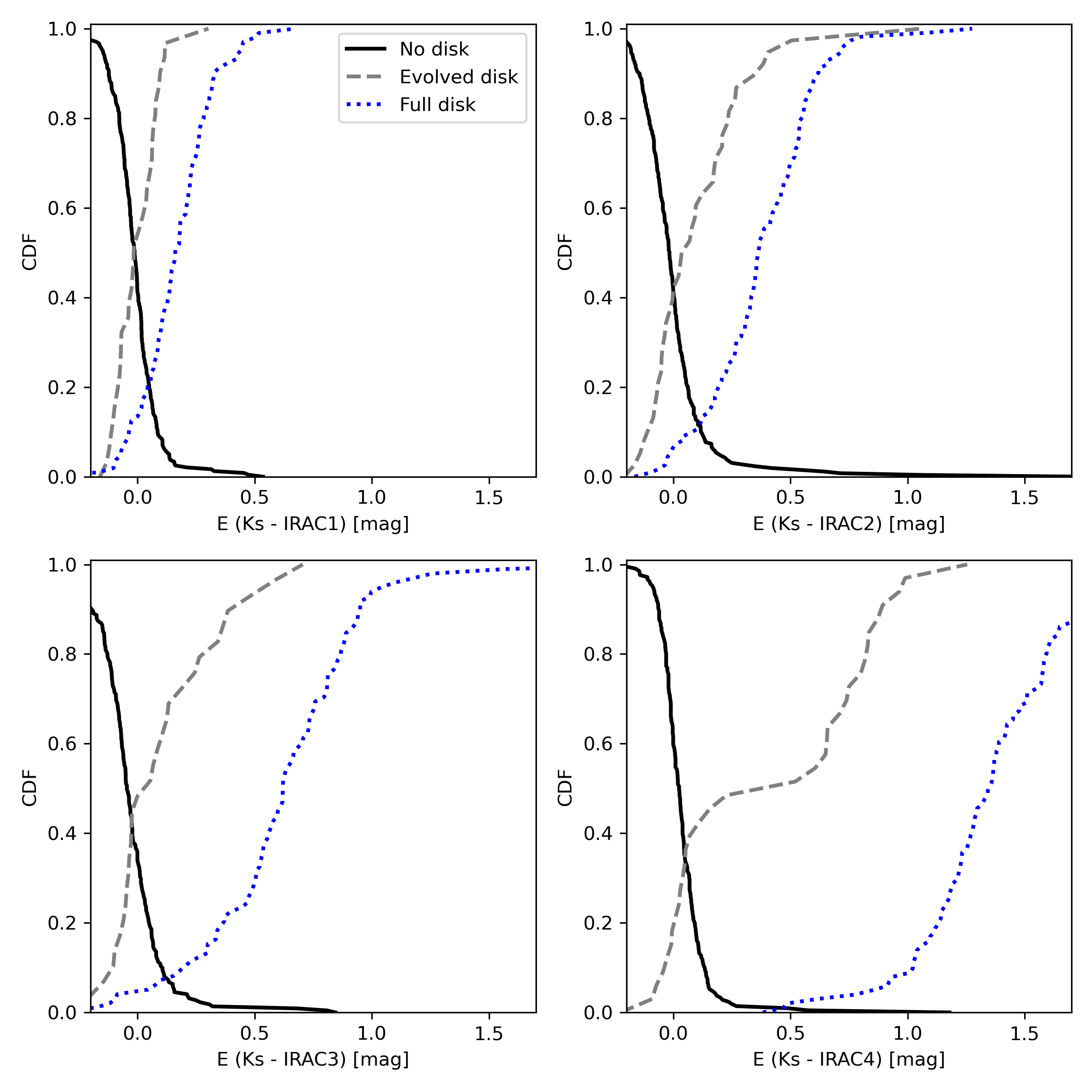}
    \caption{CDF of the $K_{\mathrm{S}}$ - IRAC colors for the three disk classifications: Full (blue dotted line), evolved (grey dashed line) and diskless (black solid line, we show 1-CDF for visualization purposes).}
    \label{fig:app_ir_colors_cdf}
\end{figure*}

We also estimate what is the typical precision and completeness of the 0.25 mag excess boundary we have used to identify sources with infrared excess in Sect.~\ref{analysis_excess}. We add all the disk-bearing sources to the compilation of diskless members of the same regions. The disk-bearing sources all have classification of the type of disk they host from the literature. We divide the disk-bearing sources in ``full'' (full and transitional disks) and ``evolved'' (debris and evolved transitional) disks. In Fig.~\ref{fig:app_ir_colors_cdf} we show the color excess CDF for the three classes disk classifications (for diskless sources we show 1-CDF for visualization purposes). In Table~\ref{tab:tab_app_irac_colors_prec} we provide the percentage of sources above the 0.25 mag color excess threshold for the sources with full disks, with evolved disks and without disks. We observe that in the $K_{\mathrm{S}}-$IRAC1 and $K_{\mathrm{S}}-$IRAC2 colors there is a significant population of ``full'' disks that are not recovered. In the four colors, using a 0.25 mag color excess for classification we have a contamination of diskless sources $<$3\% but we are missing most of the ``evolved'' disk sources as we expected.

\begin{table}
    \caption{Percentage of sources above the 0.25 mag $K_{\mathrm{S}}$ - IRAC color excess for ``full'' disk, ``evolved'' disk and no disk sources.}
    \begin{center}
        \begin{tabular}{l c c c c}
            \hline\hline
            Disk & Ks-IRAC1 & Ks-IRAC2 & Ks-IRAC3 & Ks-IRAC4 \\ 
            \hline
            No disk & 1.7 & 2.7 & 2.3 & 1.4 \\ 
            Evolved & 0.0 & 15.8 & 20.7 & 48.5 \\ 
            Full & 27.9 & 73.2 & 87.8 & 100.0 \\ 
            \hline 
        \end{tabular}
    \end{center}
    \label{tab:tab_app_irac_colors_prec}
\end{table}

\begin{table}
    \caption{$K_{\mathrm{S}}$ - IRAC photospheric colors as a function of SpT.}
    \begin{center}
        \begin{tabular}{l c c c c}
            \hline\hline
            SpT & $K_{\mathrm{S}}$-IRAC1 & $K_{\mathrm{S}}$-IRAC2 & $K_{\mathrm{S}}$-IRAC3 & $K_{\mathrm{S}}$-IRAC4 \\
              & [mag] & [mag] & [mag] & [mag] \\
            \hline
            K0 & 0.05 & 0.0 & 0.0 & 0.11 \\ 
            K0.5 & 0.05 & 0.005 & 0.005 & 0.11 \\ 
            K1 & 0.05 & 0.01 & 0.011 & 0.11 \\ 
            K1.5 & 0.05 & 0.015 & 0.016 & 0.11 \\ 
            K2 & 0.05 & 0.02 & 0.021 & 0.11 \\ 
            K2.5 & 0.05 & 0.025 & 0.027 & 0.11 \\ 
            K3 & 0.05 & 0.03 & 0.032 & 0.16 \\ 
            K3.5 & 0.05 & 0.035 & 0.037 & 0.164 \\ 
            K4 & 0.05 & 0.04 & 0.043 & 0.185 \\ 
            K4.5 & 0.05 & 0.045 & 0.048 & 0.2 \\ 
            K5 & 0.05 & 0.05 & 0.053 & 0.21 \\ 
            K5.5 & 0.05 & 0.055 & 0.059 & 0.22 \\ 
            K6 & 0.071 & 0.06 & 0.064 & 0.215 \\ 
            K6.5 & 0.065 & 0.065 & 0.069 & 0.219 \\ 
            K7 & 0.08 & 0.07 & 0.075 & 0.217 \\ 
            K7.5 & 0.095 & 0.075 & 0.08 & 0.218 \\ 
            K8 & 0.095 & 0.08 & 0.085 & 0.22 \\ 
            K8.5 & 0.095 & 0.085 & 0.131 & 0.222 \\ 
            K9 & 0.095 & 0.09 & 0.138 & 0.225 \\ 
            K9.5 & 0.095 & 0.095 & 0.144 & 0.228 \\ 
            M0 & 0.095 & 0.1 & 0.15 & 0.23 \\ 
            M0.5 & 0.095 & 0.105 & 0.156 & 0.265 \\ 
            M1 & 0.062 & 0.118 & 0.17 & 0.27 \\ 
            M1.5 & 0.074 & 0.147 & 0.175 & 0.28 \\ 
            M2 & 0.095 & 0.158 & 0.21 & 0.292 \\ 
            M2.5 & 0.129 & 0.183 & 0.223 & 0.33 \\ 
            M3 & 0.133 & 0.184 & 0.218 & 0.345 \\ 
            M3.5 & 0.167 & 0.206 & 0.216 & 0.36 \\ 
            M4 & 0.193 & 0.243 & 0.244 & 0.392 \\ 
            M4.5 & 0.211 & 0.27 & 0.267 & 0.431 \\ 
            M5 & 0.245 & 0.312 & 0.325 & 0.49 \\ 
            M5.5 & 0.28 & 0.361 & 0.385 & 0.534 \\ 
            M6 & 0.32 & 0.422 & 0.428 & 0.596 \\ 
            M6.5 & 0.341 & 0.45 & 0.471 & 0.62 \\ 
            M7 & 0.391 & 0.498 & 0.496 & 0.62 \\ 
            M7.5 & 0.415 & 0.533 & 0.56 & 0.66 \\ 
            M8 & 0.45 & 0.545 & 0.568 & 0.734 \\ 
            M8.5 & 0.48 & 0.57 & 0.58 & 0.76 \\ 
            M9 & 0.595 & 0.703 & 0.801 & 0.825 \\ 
            M9.5 & 0.67 & 0.8 & 0.965 & 1.16 \\ 
            L0 & 0.705 & 0.807 & 0.968 & 1.202 \\ 
            \hline 
        \end{tabular}
    \end{center}
    \label{tab:tab_app_irac_colors}
\end{table}

\section{SpT - $T_{\mathrm{eff}}$ scales}
\label{app_teff_scale}

\begin{figure*}[hbt!]
    \centering
    \includegraphics[width=\textwidth]{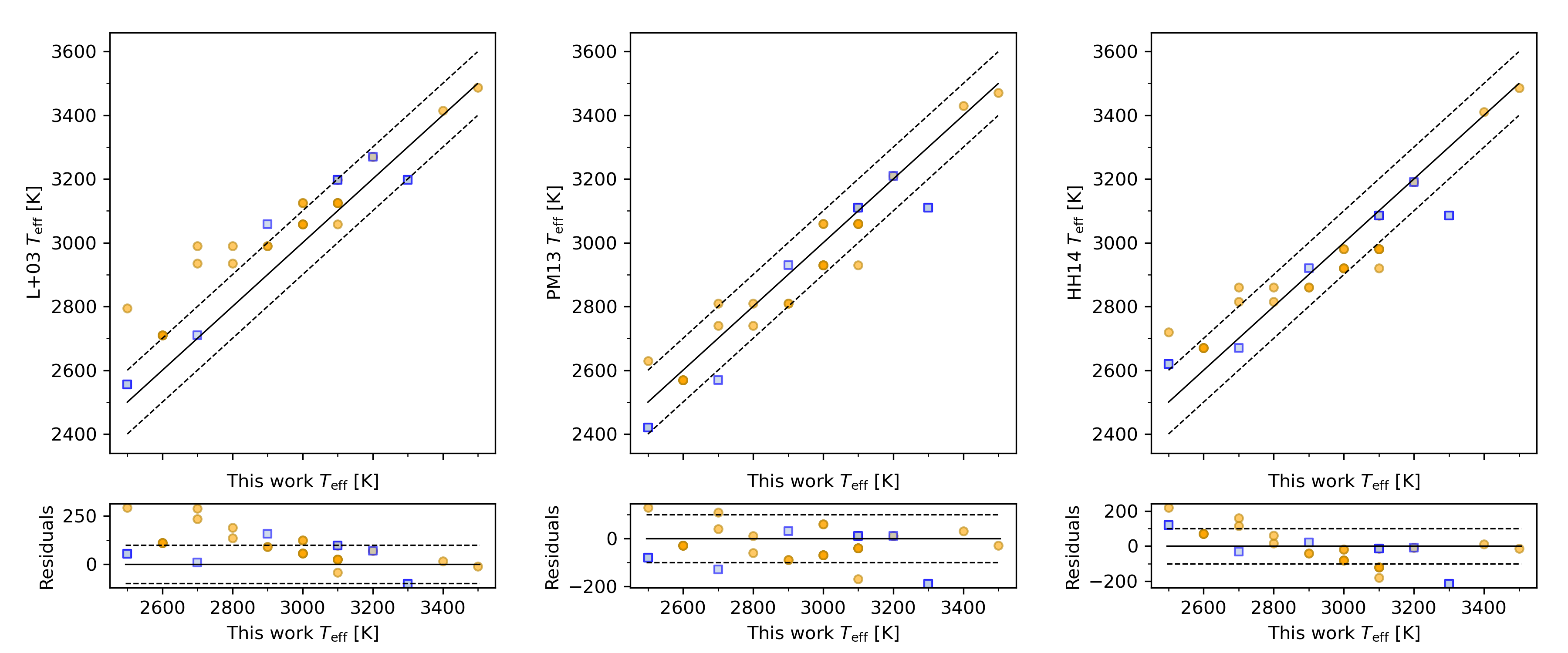}
    \caption{Comparison of the $T_{\mathrm{eff}}$ derived in this work for the KMOS spectra (see Sect.~\ref{analysis_hrd}) compared with the  $T_{\mathrm{eff}}$ derived using the scales from \citet[L03]{luhman03a}, \citet[PM13]{pecaut13} and \citet[HH14]{herczeg14}. The black line represents the linear relationship between the variables and the black dashed lines represent a $\pm$100 K uncertainty.}
    \label{fig:app_teff_compare}
\end{figure*}

In Fig.~\ref{fig:app_teff_compare} we show the comparison between the $T_{\mathrm{eff}}$ we derived in Sect.~\ref{analysis_hrd} from comparing the KMOS spectra with the BT-Settl atmospheric models \citep{baraffe15} and the $T_{\mathrm{eff}}$ converted from the SpT obtained in Sect.~\ref{analysis_templates} using the SpT-$T_{\mathrm{eff}}$ relationships of \citet{luhman03a}, \citet{pecaut13} and \citet{herczeg14}. We observe that while the three scales agree well with our $T_{\mathrm{eff}}$ there are certain important aspects to highlight. The scale from \citet[left panel]{luhman03a} systematically produces larger $T_{\mathrm{eff}}$ than what we find by fitting the spectra with the BT-Settl models. This effect is specially important above M5.5 ($\sim$3000 K), which could be associated with the dramatic evolution of atmospheric models since this scale was defined. The \citet[right panel]{herczeg14} scale has good correlation with our $T_{\mathrm{eff}}$, but from M6 ($\sim$2800) it also provides larger $T_{\mathrm{eff}}$. The scale from \citet[middle panel]{pecaut13} agrees best with our $T_{\mathrm{eff}}$ values despite having been defined for field objects. In order to be as consistent as possible, we have used the SpT-$T_{\mathrm{eff}}$ relationship from \citet{pecaut13} to obtain $T_{\mathrm{eff}}$ for the sources with only SpT available, since we have found it to perform most similarly with the $T_{\mathrm{eff}}$ we have measured by comparing the KMOS spectra with the BT-Settl atmospheric models.

\section{NGC 2244 members}
\label{app_all_members}

In Table~\ref{tab:tab_app_members} we provide the ID, coordinates, SpT, $A\mathrm{_V}$, $T\mathrm{_{eff}}$, whether they have KMOS or VIMOS spectrum (Obs), membership assessment from \citet{muzic22} and \citet{meng17}, the infrared excess key derived in Sect.~\ref{analysis_excess} and the final membership assessment for all the members and candidate members used in this work (see Sect.~\ref{analysis_census}).

\begin{table*}
    \caption{Membership analysis of all the members and candidate members of NGC 2244 used in this work. Only the first 10 lines are shown, this table will be made available in its entirety.}
    \begin{center}
        \begin{tabular}{r c c c c c c c c c c c c c}
            \hline\hline
            ID & RA & DEC & SpT & $A_{\mathrm{V}}$ & $T_{\mathrm{eff}}$ & \multicolumn{2}{c}{KMOS} & \multicolumn{2}{c}{VIMOS} & M22\tablefootmark{a}  & M17\tablefootmark{b}  & Excess\tablefootmark{c}  & Member \\ 
             & [º] & [º] &  & [mag] & [K] & Obs & Ind\tablefootmark{d}  & Obs & H$\alpha$\tablefootmark{e}  &   &   &   &    \\ 
            \hline
            3 & 97.961 & 4.97661 & M1.75 & 2.4 & 3590 & N & - & Y & N & N & Y & N & ? \\ 
            5 & 97.96189 & 4.96056 & M5.5 & 6.6 & 3100 & Y & Y & N & - & - & - & Y & Y \\ 
            6 & 97.96191 & 4.95555 & M6.0 & 3.0 & 2900 & Y & Y & N & - & - & - & Y & Y \\ 
            18 & 97.96784 & 4.97609 & M5.0 & 0.8 & 3100 & Y & Y & Y & N & N & Y & N & Y \\ 
            25 & 97.97005 & 4.96949 & M6.25 & 0.4 & 2775 & N & - & Y & Y & N & Y & N & Y \\ 
            31 & 97.97226 & 4.95376 & - & 0.25 & 2600 & N & - & N & - & Y & - & Y & Y \\ 
            32 & 97.97254 & 4.94106 & M1.75 & 1.8 & 3500 & N & - & Y & N & Y & Y & N & Y \\ 
            33 & 97.97304 & 4.94408 & - & 3.25 & 4600 & N & - & N & - & Y & Y & Y & Y \\ 
            34 & 97.97431 & 4.97966 & M3.5 & 0.8 & 3270 & N & - & Y & N & N & Y & N & ? \\ 
            35 & 97.97451 & 4.95315 & - & 3.25 & 5000 & N & - & N & - & Y & - & Y & Y \\ 
            \hline 
        \end{tabular}
        \\$^a$Membership in \citet{muzic22}.
        $^b$Membership in \citet{meng17}.
        $^c$Infrared excess key derived in Sect.~\ref{analysis_excess}.
        $^d$KMOS gravity-sensitive spectral indices key defined in Sect.~\ref{analysis_youth}.
        $^e$VIMOS H$\alpha$ emission consistent with accretion.
    \end{center}
    \label{tab:tab_app_members}
\end{table*}

\end{appendix}

%
%

\end{document}